\documentclass{article} 
\usepackage{collas2025_conference,times}
\usepackage{easyReview}
\usepackage{algcompatible}
\usepackage{algpseudocode}
\usepackage[linesnumbered,ruled,vlined]{algorithm2e}
\usepackage{graphicx}
\usepackage{amsmath}
\usepackage{amssymb}
\usepackage{booktabs}
\usepackage{tabu}
\usepackage{pifont}
\usepackage{xcolor}
\usepackage{subcaption}
\usepackage{wrapfig}
\usepackage{graphicx}
\usepackage{arydshln}
\makeatletter
\def\adl@drawiv#1#2#3{%
        \hskip.5\tabcolsep
        \xleaders#3{#2.5\@tempdimb #1{1}#2.5\@tempdimb}%
                #2\z@ plus1fil minus1fil\relax
        \hskip.5\tabcolsep}
\newcommand{\cdashlinelr}[1]{%
  \noalign{\vskip\aboverulesep
           \global\let\@dashdrawstore\adl@draw
           \global\let\adl@draw\adl@drawiv}
  \cdashline{#1}
  \noalign{\global\let\adl@draw\@dashdrawstore
           \vskip\belowrulesep}}
\makeatother

\newcommand{\minisection}[1]{\vspace{0.005in} \noindent {\bf #1}}
\newcommand{\Base}{\textit{\textsc{base}}\@}
\newcommand{\Bait}{\textit{\textsc{bait}}\@}
\newcommand{\Clean}{\textit{\textsc{clean}}\@}
\newcommand{\Multibase}{\textit{\textsc{multibase}}\@}
\newcommand{\Multibait}{\textit{\textsc{multibait}}\@}

\usepackage{amsmath,amsfonts,bm}

\def\eqref#1{equation~\ref{#1}}

\def\1{\bm{1}}

\DeclareMathAlphabet{\mathsfit}{\encodingdefault}{\sfdefault}{m}{sl}
\SetMathAlphabet{\mathsfit}{bold}{\encodingdefault}{\sfdefault}{bx}{n}

\usepackage{hyperref}
\hypersetup{
    colorlinks=true,
    linkcolor=red,
    filecolor=magenta,
    urlcolor=blue,
    citecolor=purple,
    pdftitle={Overleaf Example},
    pdfpagemode=FullScreen,
    }

\title{Addressing the Devastating Effects of Single-Task Data Poisoning in Exemplar-free Continual Learning}

\author{
  Stanisław Pawlak$^{1}$\thanks{Corresponding author: stanislaw.pawlak.dokt@pw.edu.pl} 
  \quad\quad
  Bartłomiej Twardowski$^{2,3}$ 
    \quad\quad
  Tomasz Trzciński$^{1,2}$ 
  \quad\quad
  Joost van de Weijer$^{3}$ 
  \\
  \\
  $^1$ Warsaw University of Technology, Poland 
  \quad
  $^2$ IDEAS Research Institute, Poland 
  \\
  $^3$ Computer Vision Center, Universitat Autonoma de Barcelona, Spain
}
\preprintcopy 

\begin{document}

\maketitle

\begin{abstract}
Our research addresses the overlooked security concerns related to data poisoning in continual learning (CL). Data poisoning – the intentional manipulation of training data to affect the predictions of machine learning models – was recently shown to be a threat to CL training stability. While existing literature predominantly addresses scenario-dependent attacks, we propose to focus on a more simple and realistic single-task poison (STP) threats. In contrast to previously proposed poisoning settings, in STP adversaries lack knowledge and access to the model, as well as to both previous and future tasks. During an attack, they only have access to the current task within the data stream. Our study demonstrates that even within these stringent conditions, adversaries can compromise model performance using standard image corruptions. We show that STP attacks are able to strongly disrupt the whole continual training process: decreasing both the stability (its performance on past tasks) and plasticity (capacity to adapt to new tasks) of the algorithm. Finally, we propose a high-level defense framework for CL along with a poison task detection method based on task vectors. The code is available at: \url{https://github.com/stapaw/STP.git}.
\end{abstract}

\section{Introduction}
\label{sec:intro}

Continual learning (CL) aims to accumulate knowledge from a stream of data, allowing models to adapt to new information while preventing forgetting of already acquired knowledge~(\cite{delange2021continual,masana2022class}). This continuous process seeks to strike a balance between stability and plasticity, enabling the continual learner to maintain performance on previous data while adapting to new data. CL training is based on the implicit assumption that a stream of training data is representative of the data encountered during inference time -- samples are from the same domain or share similar characteristics. Data poisoning attacks violate this assumption by manipulation of the training data aimed at degradation of model performance at inference time~(\cite{Cin__2023}). 

Recent works~(\cite{li2022targeted, umer2020targeted, pmlr-v202-kang23c, Abbasi_2024_CVPR, han2023data}) demonstrate successful data poisoning attacks on CL, highlighting the importance of the security challenges present in various CL environments. Attacks against CL proposed so far differ in terms of considered threat model, learning scenarios and poisoning methods. With so many hidden assumptions, it is much harder to develop and propose adequate defenses. Moreover, most attacks use a non-restrictive threat model which is not realistic: adversaries have access to multiple tasks within the learning stream~(\cite{umer2020targeted,li2022targeted, li2023pacol, han2023data}) and/or have the ability to access the trained model during training to produce adversarial samples as a poison~(\cite{li2022targeted, li2023pacol, han2023data}).

\begin{figure}[t]
  \centering
  \includegraphics[width=0.7\textwidth]{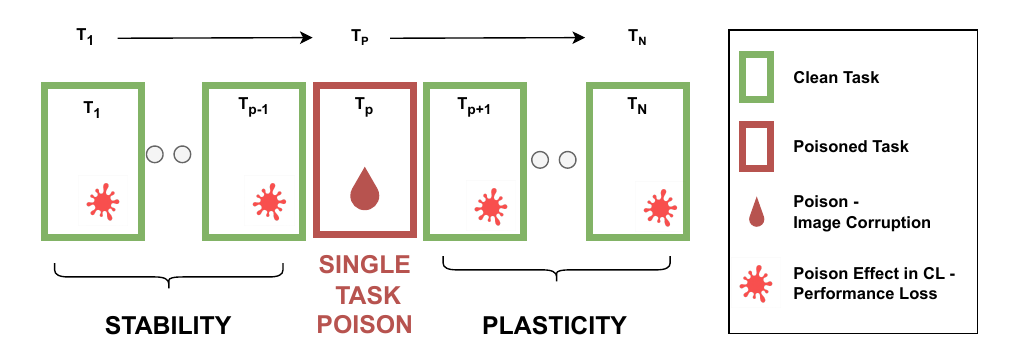} 
  
  \caption{\textbf{The Single-Task Poison (STP) setting.} We propose a new, more realistic CL setup to investigate data poisoning attacks, where an adversary has only access to a single task ($T_p$) with no knowledge about other tasks in the sequence and no access to the trained model. 
  We show that STP attacks can strongly affect the stability and the plasticity of the model.
  }
   \label{fig:STP-teaser}
   \vspace{-0.5cm}
\end{figure}

To address these issues, we propose a new, more simple and realistic Single-Task Poisoning (STP) setup where an adversary has only access to a current CL task data, and access to the trained model is not allowed. To assess how much damage could be done when poisoning a single task, we propose two types of poisoning attacks: uniform and classifier bait. The former contaminates the task data with one of the standard image corruptions (see Figure~\ref{fig:STP-corruptions-severities}). The latter corrupts only a subset of classes within a task to create data distribution differences between classes and uses corruptions as a bait for the classifier to focus on those artificially introduced spurious correlations rather than real class-discriminatory features. We want to highlight, that the goal of the STP framework is not to assess the damage done to the poisoned task performance, but rather to investigate the performance on previous and future tasks in the sequence during the CL training session. The common aim of the attacks proposed so far was to affect the stability of the model by increasing the forgetting on previous tasks. 
To the best of our knowledge, we are the first to investigate how data poisoning attacks can affect model performance on future tasks (plasticity) and the first to show an attack that is affecting both the stability and plasticity of the model at the same time.

We summarise the contributions of this work as follows:
\begin{itemize}
    \item We focus on class-incremental setup with exemplar-free methods and propose the Single-Task Poison~--~a~more realistic adversarial setup for CL, where adversary' data access is restricted only to one task in the sequence, with no knowledge about previous tasks or access to the trained model.
    \item We show that even under this restricted threat model, data poisoning attacks can have devastating effects. To the best of our knowledge, we are the first to investigate the consequences of poisoning on future tasks performance. We show that an attack can increase catastrophic forgetting and impede future training at the same time.
    \item We present a high-level defense framework for CL with a poison task detection method based on task vectors.
\end{itemize}

\section{Related work}

\minisection{Data poisoning attacks.} Data poisoning attacks are considered one of the strongest concerns for the use of offline-trained ML models in real life~(\cite{Cin__2023}), and thus multiple attacks~(\cite{nguyen2020input, yang2017generative, munoz2017towards, shafahi2018poison}) and defenses~(\cite{taheri2020defending, hong2020effectiveness, hayase2021spectre,zhu2021clear, yang2022not}) were proposed. The main three categories of data poisoning attacks are~(\cite{Cin__2023}): indiscriminate, targeted and backdoor attacks. 1)~Indiscriminate attacks involve poisoning the training data with the intention of compromising the overall performance of the model. They introduce noise or biases in a way that undermines the model's accuracy and generalization. 2)~Targeted attacks aim to manipulate the model's behavior in particular situations or for certain inputs.  Poisoned data influence the model's predictions for certain targeted samples while maintaining normal performance for all the others. 3)~Backdoor attacks are targeted attacks that poison the data with a hidden pattern (trigger) added to training samples. This trigger does not impede network performance during training, but allow adversaries to manipulate the model's behavior during inference time. The backdoor remains hidden and passive under normal circumstances but can be activated by adding a trigger to specific samples during inference time. Using triggers adversaries can control the model's outputs in a predefined manner (e.g. by changing the prediction of sample with backdoor to a specific label).

\begin{table}[ht!]
\centering
\caption{\textbf{Comparison of data poisoning attacks in CL.} CL scenarios: DI -- Domain Incremental, TI -- Task Incremental, CI -- Class Incremental~(\cite{vandeven2019scenarios}). Data Access: assumption of the adversary' access to a stream of data. Model Access: assumption of the adversary' access to the continually trained model. Our approach is the first data poisoning attack considering CI with single access to the training data stream and without access to the trained model.}
\label{tab:related_work}
\vspace{-0.4cm}
\begin{tabular}{lccccc}
\hline
\multicolumn{4}{l}{\textbf{CL data poisoning attacks}} & \multicolumn{2}{c}{\textbf{Threat Model}} \\
\hline
Work & \multicolumn{3}{c}{CL Scenario} & {Data} & {Model} \\
& DI & TI & CI & Access &  Access\\

\hline
\cite{umer2020targeted} & \ding{55} & \ding{51} & \ding{55} &  Multiple  & \ding{55}\\
\cite{li2022targeted} & \ding{55} & \ding{51} & \ding{55} & Multiple & \ding{51} \\
\cite{li2023pacol} & \ding{51} & \ding{51} & \ding{55} & Multiple & \ding{51} \\
\cite{han2023data} & \ding{55} & \ding{51} & \ding{55} & Multiple & \ding{51}\\
\cite{pmlr-v202-kang23c} & \ding{55} & \ding{51} & \ding{55} & Single & \ding{55} \\
\cite{Abbasi_2024_CVPR}  & \ding{55} & \ding{51} & \ding{55}  & Single & \ding{51} \\
\hline

Ours & \ding{55} & \ding{55} & \ding{51} & Single & \ding{55} \\
\hline
\end{tabular}
\vspace{-0.5cm}

\end{table}

The issue of data poisoning in CL setups is largely unexplored, with only a few recent works discussing specific attacks~(\cite{li2022targeted, umer2020targeted, pmlr-v202-kang23c, Abbasi_2024_CVPR}) and defenses~(\cite{umer2023adversary, wang2022towards}). Table~\ref{tab:related_work} compares data poisoning methods in CL. Most focus on Task Incremental scenarios, where task IDs are available during inference, while our work emphasizes the understudied Class Incremental scenario. Backdoor attack methods~(\cite{umer2020targeted, umer2022false}) require data access during training and inference to trigger the attack. The most common attack technique relies on accessing the trained model to create adversarial samples for previous tasks~(\cite{li2022targeted, li2023pacol, han2023data}) or to revert the model for task information~(\cite{Abbasi_2024_CVPR}). We share a similar goal with~\cite{pmlr-v202-kang23c} in making the threat model more realistic. However, while~\cite{pmlr-v202-kang23c} focuses on exploiting vulnerabilities in generative replay (a generative model is unable to generate a label-changing pattern that is present during first classifier training, but is not generated during replay phase, which leads to a postponed label-switching attack and increase in the forgetting), our focus differs.  

\minisection{Data poisoning defenses}
Most of the standard defenses against data poisoning (like training data sanitization and outlier detection methods) assume that only a small fraction of the training data is poisoned~\cite{Cin__2023}. Although this assumption is correct for a standard stationary training, we argue that in CL this assumption is not certain. We show how an adversary can even fully poison a small task in a CL sequence, without being noticed immediately (see~\ref{fig:val_test_evaluation_defense}). Thus, most standard defenses against data poisoning are not suited to CL training. Although data poisoning in CL have recently gained increasing attention resulting in multiple new attacks proposed, there is a significant gap in terms of discussing and proposing effective defensive measures. Defensive approaches were recently proposed only for a specific subset of poisoning attacks, namely backdoor attacks (\cite{umer2023adversary, wang2022towards}).
A subset of methods based on robust training, model inspection and model sanitization (\cite{Cin__2023}) could be adapted to prevent data poisoning in CL, but as of now they are still not evaluated under CL setup and thus there is no off-the-shelf defense yet.

\minisection{CL Threat model.} We consider the simplest and most realistic scenario of the attack: an adversary has access only to the data of one task in the sequence. To our knowledge, this is the most constrained threat model yet considered. Unlike previous models, our 1) adversary lacks access to the CL model~(\cite{li2022targeted, li2023pacol, Abbasi_2024_CVPR, han2023data}), and 2) adversary has no access to or knowledge of data from previous~(\cite{li2022targeted, li2023pacol, han2023data, wang2022towards}) or 3) future tasks~(\cite{umer2020targeted, umer2022false, umer2023adversary}).

\section{Single-Task Data Poisoning}

In this section, we present Single-Task Poisoning (STP) framework to investigate the effects of data poisoning attacks in continual learning. We start with an overview and justification of STP, followed by the description of data poisoning methods, attacks, and defenses proposed in this work.

\subsection{Single Task Data Poisoning: What It Is and Why It Matters}\label{sec:STP-overview}

STP is a setup to investigate the effects of data poisoning in CL. It puts more restrictions on the adversary, who has only access to a single task data and cannot access the trained model. STP is motivated by the fact that in real-life scenarios we have access neither to the model nor to all available training data stream. In the following, we show these constraints, explain and justify the STP setting by considering a defensive party that is continually training a model on a data stream obtained from multiple sources (see Figure~\ref{fig:STP-teaser}).

Firstly, STP employs a more restrictive threat model, because the potential adversary data provider has no access to the data of other providers, nor to the trained model, and he or she does not know when the data would be used for training. 
Secondly, STP considers poisoning high-percentages of data samples, which is not common for the data poisoning settings. In stationary training, a single data provider is not able to poison a high percentage of training dataset due to the data volume. Contrary, CL training is divided into multiple stages and poisoning heavily one small task is possible.
Finally, STP provides a simple threat model to investigate how much damage could be made by a single poisoned CL task. It is true that the data from a single provider could be used within the one task or split into several tasks which could introduce additional vulnerabilities. While many previous attacks consider multi-stage attacks, we argue to focus firstly on a simpler scenario. STP provides a framework to investigate the devastating effects of poison from a single task and is a step forward to build defensive mechanisms that could be used for more secure CL training in general.
  
\begin{wrapfigure}{r}{0.45\textwidth}
  \centering
  \vspace{-2em} 
  \includegraphics[width=0.48\textwidth]{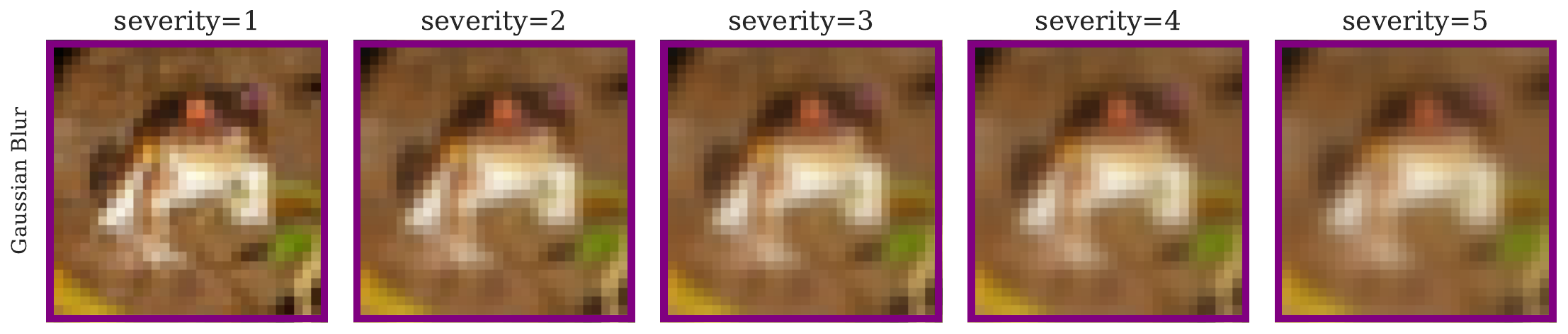}
  \caption{\textbf{Image corruption example:} Gaussian blur with increasing severity (1–5).}
  \label{fig:STP-corruptions-severities}
\end{wrapfigure}
\subsection{Data poisoning continual learning with STP}\label{sec:STP-instantiation}

\subsubsection{Task poisoning with image corruptions.}\label{sec:STP-corruptions}

We use the CIFAR-C~(\cite{hendrycks2019robustness}) set of image corruptions to transform images of the poisoned task. These corruptions are well established and commonly used, with clear publicly available algorithm and code to create them. We use them to propose the simplest possible poisoning attack, without any additional knowledge, as the STP threat model is very restrictive towards the attacker and he or she cannot use other poisoning methods proposed so far (they rely on the access to the model or information about other tasks in the sequence for optimization). Additionally, proposed different severity levels were used in the experiments, to show the strength vs. detectability trade-off faced by the attacker~\cite{frederickson2018attack}.

While CIFAR-C dataset is usually used for test-time adaptation scenarios and thus benchmark consists of corrupted test data, we use those predefined transformations to create corrupted images for training the poisoned task. Adapting these predefined transformations results in a clean-label, non-adversarially optimized data poisoning (poisoning based on a pixel distribution change after transformations like e.g. gaussian blur). For examples of corruption severities, see Figure~\ref{fig:STP-corruptions-severities}. All corruption types are shown in the Appendix~\ref{fig:STP-corruptions}.

Let $T_{p}$ be the be the task that is poisoned during CL  training with $D_{p}$ being the dataset used in that task. Let $D_{clean}$ be the original dataset for the attacked task, $D_{poisoned}$ be the poisoned copy of $D_{clean}$.

\begin{wrapfigure}{l}{0.5\textwidth}
  \centering
  \vspace{-1em} 
  \includegraphics[width=0.5\textwidth]{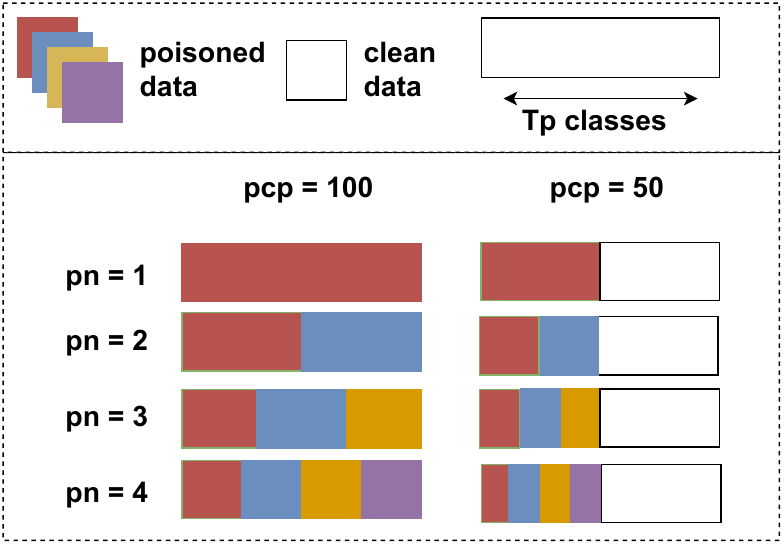}
  \caption{\textbf{Main parameters of STP attacks:} $pcp$ defines how many classes are poisoned in the poisoned task $T_p$, $pn$ how many poisons (corruptions) are used.}
  \label{fig:STP-attacks}
  \vspace{-3em}
\end{wrapfigure}

Transformation from $D_{clean}$ into $D_{poisoned}$ is defined by three parameters (see Figure~\ref{fig:STP-attacks}):
\begin{itemize}
    \item $pcp$ -- poisoned class percentage, defining how many classes are affected by corruptions in $D_{poisoned}$
    \item $pn$ -- number of poisons (i.e. different corruptions) used to poison data in $D_{poisoned}$
    \item $pp$ -- the percentage of exemplars poisoned (corrupted) in each class of $D_{poisoned}$
\end{itemize}

The poisoning attack is changing the distribution of $T_{p}$ data by applying one of the corruptions to $pp$ \% of images from $pcp$ classes of $D_{clean}$. Let $I_n$ be the subset of $n$ indices randomly chosen from $D_{clean}$, $I_{n} \subseteq \{1, 2, \ldots, |D_{clean}|\} = I$. Then $D_{p}$ is a set containing $n$ poisoned and $|D_{clean}| - n$ clean samples:
\begin{equation}
    D_{p} = D_{poisoned}[I_{n}] \cup D_{clean}[I / I_{n}]
    \label{eq:1}
\end{equation}


\subsubsection{Attacks.}\label{sec:STP-attacks}
\textbf{Naming convention for attack evaluation.}
    
We evaluate STP with four kinds of attacks:
\begin{enumerate}
    \item \Base~-- the simplest attack with single corruption added to all classes in $T_p$,
    \item \Bait~-- use single corruption as a bait for classifier: we poison only half of the classes in $T_p$ to attract the model to discriminate rather based on introduced data distribution differences than the class differences within the same data distribution,
    \item \Multibase~-- we poison all classes in $T_p$ with different corruptions to check if the attacker can obtain additional gains by increasing the number of poisons. Classes are poisoned proportionally with $pn$ corruptions. 
    \item \Multibait~-- we poison half of the classes in $T_p$. Classes are poisoned proportionally as in \Multibase~attack.

\end{enumerate}
\newpage
\subsubsection{Defense.}\label{sec:STP-defenses}
\begin{figure}
    \centering
     \includegraphics[width=0.7\linewidth]{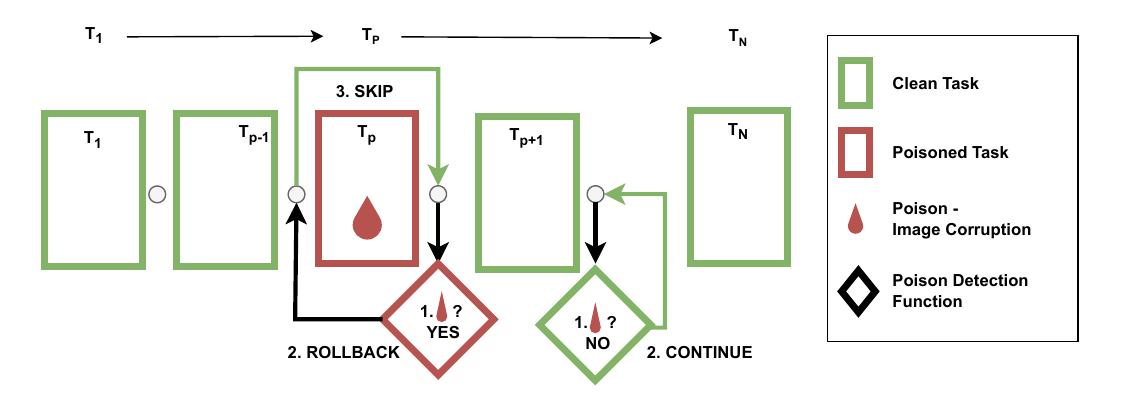}
         \caption{\textbf{The high-level defense framework.} We propose to add after each CL task a poison-detection phase to roll back changes in model weights upon detection of poison. }
               \label{fig:STP-defense}
\end{figure}
\begin{wrapfigure}{r}{0.5\textwidth}
 \vspace{-5em}
  \centering
  \includegraphics[width=\linewidth]{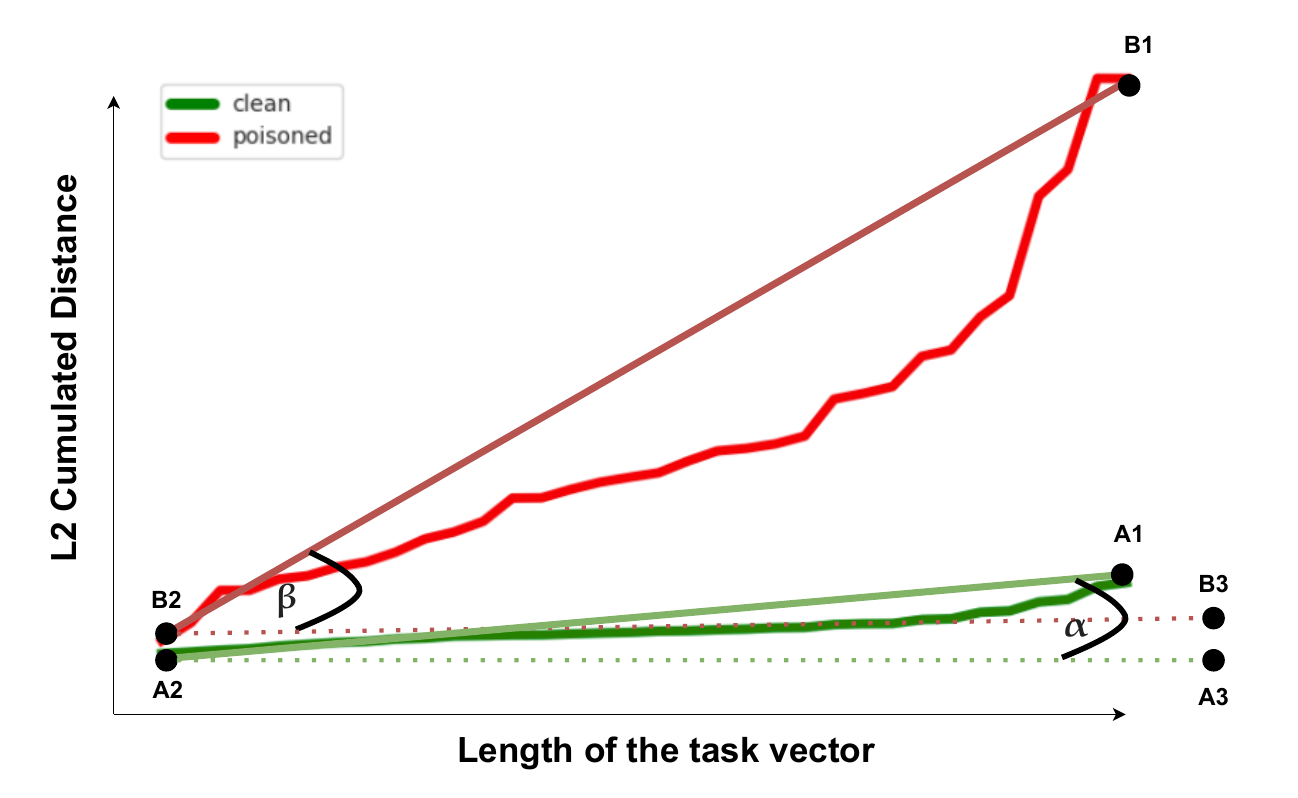}
  \caption{\textbf{Difference between clean and poisoned task vectors increases with more layers contributing to task vectors.} Consequently, we detect the poison when $\beta > \alpha$.}
  \label{fig:task-vectors-defense}
   \vspace{-1em}
\end{wrapfigure}
The defense framework involves three steps: checkpointing the model, poison task detection, and (optional) weight rollback (see Figure~\ref{fig:STP-defense}). First, we save the model weights before training on a new task. After training, a poison detection test is conducted. If the task is detected as poisoned, the model's weights are rolled back to the pre-task state, the poisoned task is skipped, and training proceeds with the next task.

\minisection{Detection using task vectors.}
The core of our framework is poisoned task detection. We propose a simple task-vector-based method that uses one clean task at the start of training to calculate a threshold. This threshold is then applied to all tasks in the stream to determine if a task is poisoned. Task vectors~(\cite{ilharco2023editingmodelstaskarithmetic})  are vector representations in the parameter space of a neural network that encapsulate the changes required to adapt the model from one task to another. They are computed by taking the difference between the model parameters fine-tuned on different tasks, effectively capturing the task-specific adaptations needed. We propose to use them for the poison task detection: by comparing task vectors for the model before and after the task training. For both clean and poisoned task distance between \textit{before} and \textit{after} vectors is increasing when we use more network layers to create task vectors, but
Figure \ref{fig:task-vectors-defense} shows that for poisoned task distance increase is more significant. Thus, we define the angle $\alpha$ with single clean task data to denote how fast the distance increase when trained on clean data (see algorithm~\ref{algorithm_alpha}). Then, we test the $\beta$ obtained for each of the following tasks. If $\beta > \alpha$ we denote the task as poisoned. 

\begin{algorithm}
\caption{Calculation of Threshold Angle $\alpha$}
\label{algorithm_alpha}
\KwIn{A sequence of tasks $\{T_1, T_2, \ldots, T_n\}$, continually trained model $M_{\theta_i}$, where $\theta_i$ are the weights of the model after training on task $T_i$}
\KwOut{Threshold angle $\alpha$}
Select a task $T_c$ from the beginning of the sequence (assumed to be clean)\;
\ForEach{seed $s_i \in \{s_1, s_2, \ldots, s_k\}$}{
    Train $M_{\text{before}}$ ($M_{\theta_{c-1}}$) on $T_c$ to obtain $M_{\text{after}}$\;
    Compute activations $A_{\text{before}}'$ and $A_{\text{after}}'$ using $T_c$ data\;
    Aggregate activations using mean into single vectors $A_{\text{before}}$ and $A_{\text{after}}$\;
    Compute task vector: $v = \|A_{\text{after}} - A_{\text{before}}\|_2$\;
    Compute cumulative sum vector: $c = \text{cumsum}(v)$\;
    Let $A_1 = c_{\text{last}},\ A_2 = c_{\text{first}}$\;
    Define $A_3$ with same x-coordinate as $A_1$ and y-coordinate as $A_2$\;
    Compute angle $\alpha'_i = \angle A_1A_2A_3$\;
}
Aggregate all angles: $\alpha = \text{Aggregate}(\{\alpha'_1, \alpha'_2, \ldots, \alpha'_k\})$ \tcp*[r]{e.g., 90th percentile or MAX}
\Return $\alpha$\;
\end{algorithm}

\section{Experiments}

\subsection{Experimental setup}
\minisection{CL setup.} In the main experiments, we use a class-incremental setup~(\cite{vandeven2019scenarios}) that divides datasets into tasks with disjoint classes and lacks task identifiers during inference. We investigate data poisoning attacks on classical exemplar-free CL methods, LwF~(\cite{li2017learning}) and EWC~(\cite{kirkpatrick2017overcoming}). Basic experiments use CIFAR10 split into five tasks of two classes each. For most experiments, we use CIFAR100, split into six tasks, with the first task having 50 classes and the others 10 classes each, to assess attacks on tasks with more than two classes.

\minisection{Model and training details.} We use the Resnet32 model for all experiments in the main section of the work. We use Resnet18 for additional experiments on Tiny-Imagenet (\cite{le2015tiny}), added in the appendix (see~\ref{attack_additional_experiments}). We use implementations from FACIL~(\cite{masana2022class}) framework for CL training. For CIFAR10 we start each task with learning rate equal to 0.05 with weight-decay 0.0001 and momentum 0.9. We train 200 epochs for each task. Additional information about learning parameters are in the appendix~\ref{setup_details}.

\minisection{Datasets} We test data poisoning attacks on CIFAR10 and CIFAR100 datasets. Both contain 32x32 images~(\cite{alex2009learning}). Additional experiments with bigger images on Tiny-Imagenet (64x64)~(\cite{le2015tiny}) are in the appendix~\ref{attack_additional_experiments}.

\minisection{Defense evaluation.} To evaluate proposed poison task detection method, we prepared a dataset of tasks created from different subsets of CIFAR100, each having ten classes. In realistic scenario clean tasks should be much more frequent than poisoned ones. Thus, we used a dataset with 92\% clean and 8\% poisoned tasks.

\subsection{Attacks evaluation}

\definecolor{mygray}{gray}{0.5}
\addtolength{\tabcolsep}{-3.5pt}

\begin{table*}[t]
\begin{center}
\caption{\textbf{Average accuracy when poisoning a single Tp task in CL sequence.} Grey (\Clean) rows show results of runs without poisoning $T_p$ task, while black (\Base, \Bait, \Multibase, \Multibait) rows show results for different attacks. Difference between \Clean~and poisoned runs is shown in the parenthesis. As most data poisoning works in CL we report accuracies after the poisoned task (ACC AT POISONING TIME). Additionally we report accuracies at the end of the full CL sequence (FINAL ACC) to show the impact of poisoning for further training. Experimental setup: CIFAR100(50/10), CIFAR10(2/2), \Base: $pcp=100$, $pn=1$, $pp=100$, \Bait: $pcp=50$, $pn=1$, $pp=100$, \Base: $pcp=100$, $pn=5$, $pp=100$, \Base: $pcp=50$, $pn=5$, $pp=100$.
}
\label{tab:main_results}
\footnotesize 
\begin{sc}
\begin{tabu}{llccc|llllll}
\toprule
Row & Attack & method & Dataset & p & \multicolumn{2}{c}{Acc at poisoning time} & \multicolumn{4}{c}{Final Acc} \\
 \cmidrule(r){6-7}
 \cmidrule(r){8-11}
 \centering
 &&&&&  $T_p$ &  before $T_p$ &  $T_p$ &  before $T_p$ &  after $T_p$ & total \\
\midrule

\rowfont{\color{mygray}} 1 & \Clean & JOINT &  CIFAR10  & 3 &  95.2 & 95.6  & 93.6 & 93.4 & 92.4 & 93.1\\
2 & \Base & JOINT &  CIFAR10  & 3 & 0.2 {\tiny (-95.0)} & 96.6 {\tiny (1.0)}  & 0.1 {\tiny (-93.5)} & 94.1 {\tiny (0.7)} & 93.1 {\tiny (0.7)} & 74.9 {\tiny (-18.2)}\\ 
3 & \Bait & JOINT &  CIFAR10  & 3 &  0.3 {\tiny (-94.9)} & 96.5 {\tiny (0.9)}& 46.2 {\tiny (-47.4)} & 93.8 {\tiny (0.4)} & 92.7 {\tiny (0.3)} & 83.8 {\tiny (-9.3)}\\
\cdashlinelr{1-11}
 \rowfont{\color{mygray}} 4 & \Clean & JOINT &  CIFAR100  & 4 & 73.0  & 71.2 & 71.3 & 68.0 & 66.5 & 68.1\\
5 & \Base & JOINT &  CIFAR100   & 4 & 1.2 {\tiny (-70.0)}& 72.2  {\tiny (1.0)}& 1.0 {\tiny (-70.3)} & 69.7 {\tiny (1.7)} & 67.1 {\tiny (0.6)} & 57.4 {\tiny (-10.7)}\\ 
6 & \Bait & JOINT &  CIFAR100 & 4 &  36.8 {\tiny (-36.2)} & 71.9 {\tiny (0.7)} & 36.0 {\tiny (-35.3)} & 68.7 {\tiny (0.7)} & 66.6 {\tiny (0.1)} & 62.6 {\tiny (-5.5)} \\ 
7 & \Multibase & JOINT &  CIFAR100  & 4 & 4.5 {\tiny (-68.5)} & 72.2 {\tiny (1.0)} & 4.3 {\tiny (-67.0)} & 69.5 {\tiny (1.5)} & 66.7 {\tiny (0.2)} & 57.7 {\tiny (-10.4)}\\ 
8 & \Multibait & JOINT &  CIFAR100  & 4 & 40.3 {\tiny (-32.7)} & 71.9 {\tiny (0.7)} & 39.7 {\tiny (-31.6)} & 69.0 {\tiny (1.0)} & 66.4 {\tiny (-0.1)} & 63.3 {\tiny (-4.8)}\\ 

\cdashlinelr{1-11}
\rowfont{\color{mygray}} 9 & \Clean & LWF &  CIFAR10  & 3 & 67.2 & 71.7 & 50.2  & 62.3 & 63.9 &  60.5 \\ 	
10 & \Base & LWF &  CIFAR10 & 3 & 56.3 {\tiny (-10.9)} & 70.1 {\tiny (-1.6)} & 39.9 {\tiny (-10.3)} & 59.1 {\tiny (-3.2)} & 65.4 {\tiny (1.5)} & 57.8 {\tiny (-2.7)} \\ 
11 & \Bait & LWF &  CIFAR10 & 3 & 45.8 {\tiny (-21.4)} & 48.5 {\tiny (-23.2)} & 41.9 {\tiny (-8.3)} & 43.8 {\tiny (-18.5)} & 50.2 {\tiny (-13.7)} & 46.0 {\tiny (-14.5)}  \\ 
\cdashlinelr{1-11}
 \rowfont{\color{mygray}} 12 & \Clean & LWF &  CIFAR100  & 4 & 78.9 & 50.8 & 61.7 & 30.6 & 69.7 & 48.8\\
13 & \Base & LWF &  CIFAR100 & 4 & 64.4 {\tiny (-14.5)} & 36.3 {\tiny (-14.5)} & 45.5 {\tiny (-16.2)} & 17.3 {\tiny (-13.3)} & 68.6 {\tiny (-1.1)} & 39.1 {\tiny (-9.7)}  \\ 
14 & \Bait & LWF &  CIFAR100 & 4 & 52.2 {\tiny (-26.7)} & 41.5 {\tiny (-9.3)} & 44.3 {\tiny (-17.4)} & 23.5 {\tiny (-7.1)} & 62.3 {\tiny (-7.4)} & 39.9 {\tiny (-8.9)} \\ 
15 & \Multibase & LWF &  CIFAR100 & 4 & 37.1 {\tiny (-41.8)} & 46.8 {\tiny (-4.0)} & 28.7 {\tiny (-33.0)} & 27.4 {\tiny (-3.2)} & 66.1 {\tiny (-3.6)} & 40.5 {\tiny (-8.3)} \\
16 & \Multibait & LWF &  CIFAR100  & 4 & 54.5 {\tiny (-24.4)} & 45.7 {\tiny (-5.1)} & 47.3 {\tiny (-14.4)} & 27.1 {\tiny (-3.5)} & 65.3 {\tiny (-4.4)} & 43.2 {\tiny (-5.6)} \\
\cdashlinelr{1-11}

\rowfont{\color{mygray}} 17 & \Clean & EWC &  CIFAR10  & 3 & 34.3 & 22.2 & 20.0  & 11.6 & 24.0 &  18.2\\ 	
18 & \Base & EWC &  CIFAR10 & 3 & 19.7 {\tiny (-14.6)} & 25.7 {\tiny (3.5)} & 4.2 {\tiny (-15.8)} & 15.0 {\tiny (3.4)} & 29.5 {\tiny (5.5)} & 18.6 {\tiny (0.4)} \\ 
 19 & \Bait & EWC &  CIFAR10 & 3 & 40.6 {\tiny (6.3)} & 6.3 {\tiny (-15.9)} & 16.7 {\tiny (-3.3)} & 3.7 {\tiny (-7.9)} & 24.9 {\tiny (0.9)} & 14.8 {\tiny (-3.4)} \\ 
\cdashlinelr{1-11}
 \rowfont{\color{mygray}} 20 & \Clean & EWC &  CIFAR100 & 4 & 82.3 & 27.8 & 35.9  & 17.8 & 57.2  & 33.9\\	
21 & \Base & EWC &  CIFAR100 & 4 & 48.2 {\tiny (-34.1)} & 17.7 {\tiny (-10.1)} & 25.6 {\tiny (-10.3)} & 18.9 {\tiny (1.1)} & 57.4 {\tiny (0.2)} & 32.9 {\tiny (-1.0)}  \\ 
22 &\Bait & EWC &  CIFAR100 & 4 & 51.9 {\tiny (-30.4)} & 15.2 {\tiny (-12.6)} & 22.4 {\tiny (-13.5)} & 18.2 {\tiny (0.4)} & 56.8 {\tiny (-0.4)} & 31.8 {\tiny (-2.1)} \\
  23 & \Multibase & EWC & CIFAR100  & 4 &  37.2 {\tiny (-45.1)} & 20.1 {\tiny (-7.7)} & 13.1 {\tiny (-22.8)} & 18.0 {\tiny (0.2)} & 58.9 {\tiny (1.7)} & 30.8 {\tiny (-3.1)} \\
  24 &\Multibait & EWC & CIFAR100  & 4 & 56.6 {\tiny (-25.7)} & 20.5 {\tiny (-7.3)} & 25.1 {\tiny (-10.8)} & 17.5 {\tiny (-0.3)} & 56.6 {\tiny (-0.6)} & 31.8 {\tiny (-2.1)} \\
\bottomrule
\end{tabu}
\end{sc}
\end{center}

\end{table*}
\addtolength{\tabcolsep}{3.5pt} 

We investigate STP attacks with these questions:
\begin{enumerate}
    \item What is the difference between poisoning a single task in exemplar-free CL compared to poisoning single task in joint training?
    \item What is the decrease in the model performance after poisoning task $T_p$ with proposed STP attacks? How poisoning affects continual learner stability – performance on tasks before $T_p$ – and plasticity – performance on the following tasks?
    \item Is it better for the attacker to poison all data with the same corruption (i. e. using \Base, \Bait~attacks), or use multiple corruptions (\Multibase, \Multibait~attacks) for different classes? 
    \item How can the adversary address the strength vs detectability trade-off~(\cite{frederickson2018attack}) of the data poisoning attack? What percentage of $T_p$ data should be poisoned for STP attack to be effective?
    
\end{enumerate}
 \begin{figure}[bt!]
 \vspace{-0.3cm}
  \centering
  \includegraphics[width=0.65\textwidth]{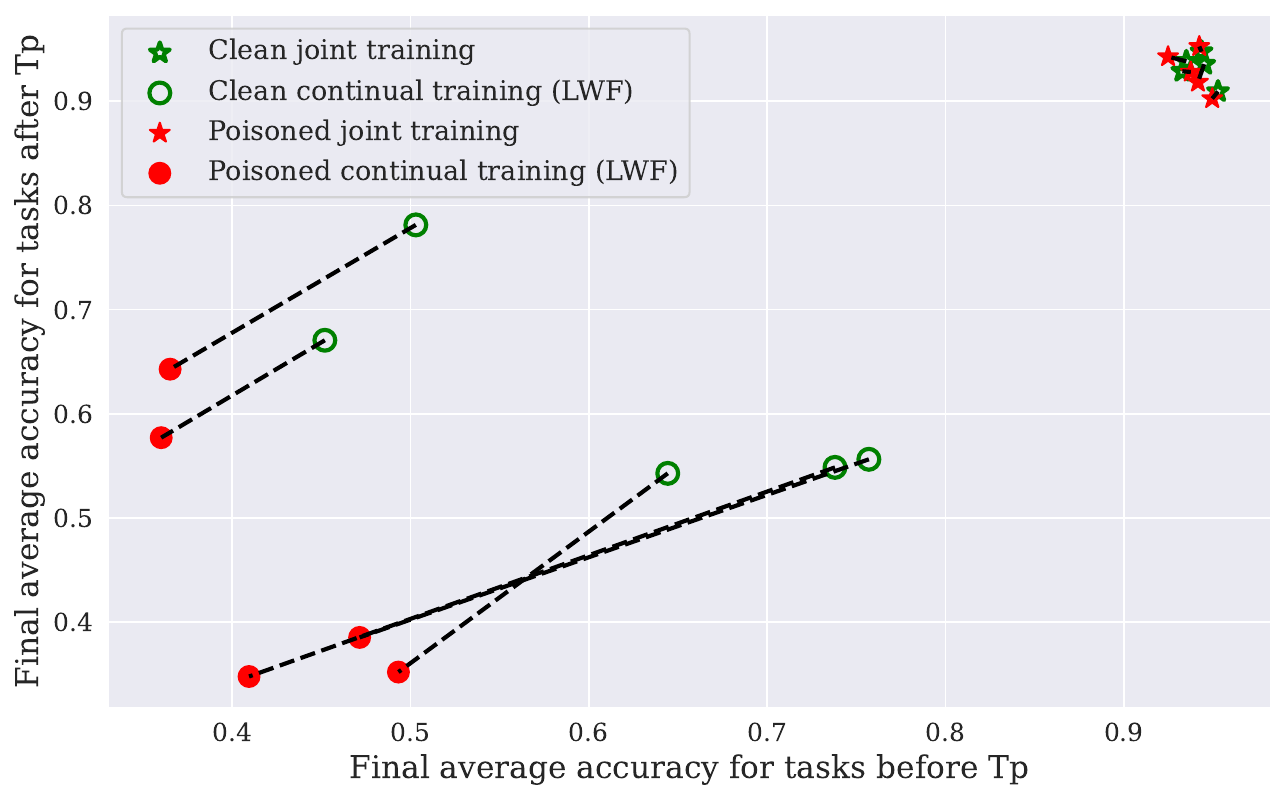} 
  
  \caption{\textbf{Poisoning single task (Tp) training data has devastating effect for the performance in CL.} In CL single task poisoning may decrease the performance for past and future classes, while in joint training, performance on non-poisoned subset of classes is not affected. Experimental setup: CIFAR10, LWF, Resnet32, $p=3$.}
   \label{fig:teaser}
   \vspace{-0.4cm}
\end{figure}
We answer them with the following observations: \newline
\minisection{1)~Poisoning has more severe consequences in CL than in joint training.} Figure~\ref{fig:teaser} shows the difference between poisoning part of the data in joint training on CIFAR10 and the same part of the data treated as single task in CL sequence. The former does not affect the performance on the rest of the classes (poisoned red-filled stars cover the same area of the plot as clean green open stars). The latter affects both the performance on classes from past and future tasks (poisoned red-filled dots are in the left-down direction from clean green-open dots). The reason behind the above observation is that in JOINT case the model is trained on clean and poisoned classes at the same time, while in CL sequence we train a model on isolated poisoned data without the access to clean samples from previously seen classes. \newline

\minisection{2A) A poisoning singe task in CL sequence decreases performance not only on $T_p$, but also on past and future tasks.} As expected, the performance decrease is usually the biggest for the poisoned task (see $T_p$ columns in Table~\ref{tab:main_results}). However, Table~\ref{tab:main_results} presents further evidence for observation \textbf{1)}: there is almost no  difference in accuracy for BEFORE $T_p$ and AFTER $T_p$ in JOINT training (for rows 2-3 and 5-8 differences between \Clean~and poisoned training in the parenthesis are close to 0), while there is a significant difference when poisoning CL sequence (for rows 10-11, 13-16 differences are negative). At the same time, we see that the relative performance drop is much bigger for tasks before $T_p$ than after: overriding patterns to distinguish old classes without access to the data is a bigger problem than struggles in learning new patterns in tasks after $T_p$. \newline

\begin{figure}
    \centering
    \begin{minipage}{0.47\textwidth}
        \centering
        \includegraphics[width=\linewidth]{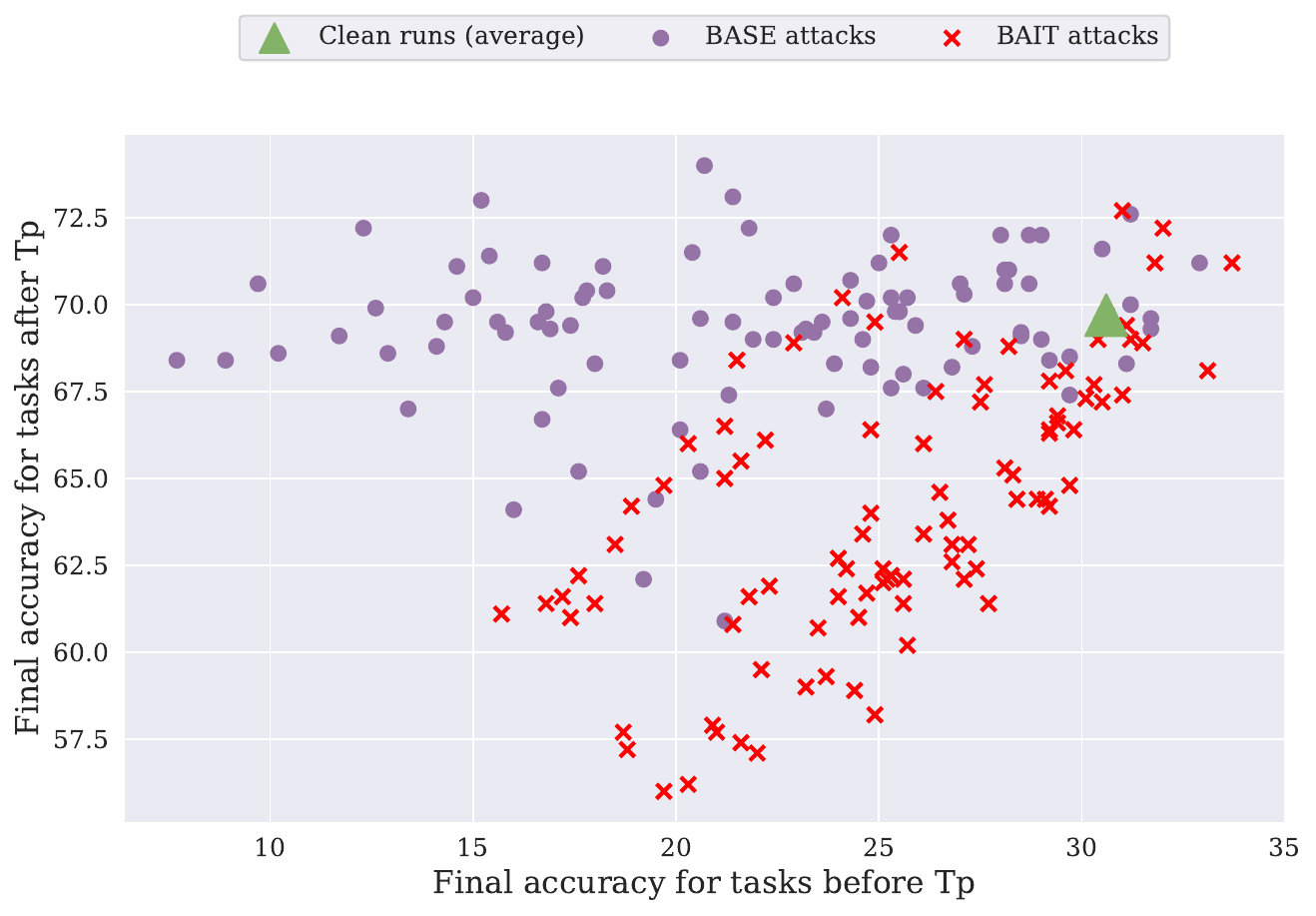}   
  \caption{\textbf{Comparison of corresponding \Base~and \Bait~attacks impact on the performance on past and future tasks.} \Bait~attacks are stronger than \Base~while poisoning much less exemplars of $T_p$. Experimental setup: CIFAR100, LWF, Resnet32.
  }
   \label{fig:base-scatter}
  
    \end{minipage}
    \hspace{0.5cm}
    \begin{minipage}{0.47\textwidth}
        \centering
        \includegraphics[width=\linewidth]{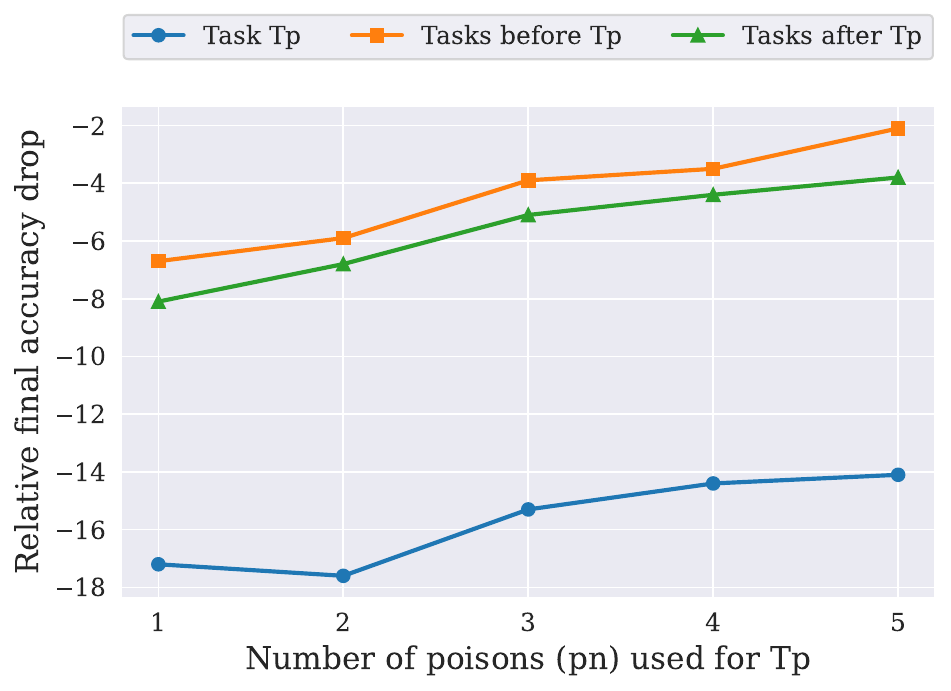}   
  \caption{\textbf{Relative accuracy drop for \Bait~($pn=1$) and~\Multibait~($pn>1$) attacks using different number of poisons.} Increasing $pn$ decrease poison effectiveness. Experimental setup: CIFAR100, LWF, Resnet32.}
   \label{fig:bait-stability}
  
    \end{minipage}
   
\end{figure}

\minisection{2B) \Bait~attacks affect performance on tasks after $T_p$ stronger than \Base~attacks.} Figure~\ref{fig:base-scatter} expands evidence presented in Table~\ref{tab:main_results} and show the effects of \Base~and \Bait~attacks with different corruptions on CIFAR100. Interestingly, \Bait~attacks are stronger than \Base~while poisoning much less exemplars of $T_p$. \Base~attacks primarily reduce accuracy on past tasks, affecting model stability, while \Bait~attacks decrease performance on both past and future tasks. We hypothesize that spurious correlations ('baits') in $T_p$ data impair future task learning more than single corruption-based data distribution shifts in \Base~attacks.\newline

\minisection{3) \Multibase~and \Multibait~attacks are less effective than their counterparts with single corruption.} 
Comparing results for multi-corruption attacks and single corruption attacks in Table~\ref{tab:main_results} we see a bigger performance decrease for the latter ones: impact of one corruption is stronger than multiple, when we poison constant number of classes (with constant $pcp$). Results presented on Figure~\ref{fig:bait-stability} support this observation. The bigger the number of used corruptions, the smaller the poisoning effect on both past and future tasks. 

\begin{figure}[bt]
    \centering
    
    \begin{subfigure}{0.47\textwidth}
        \includegraphics[width=\linewidth]{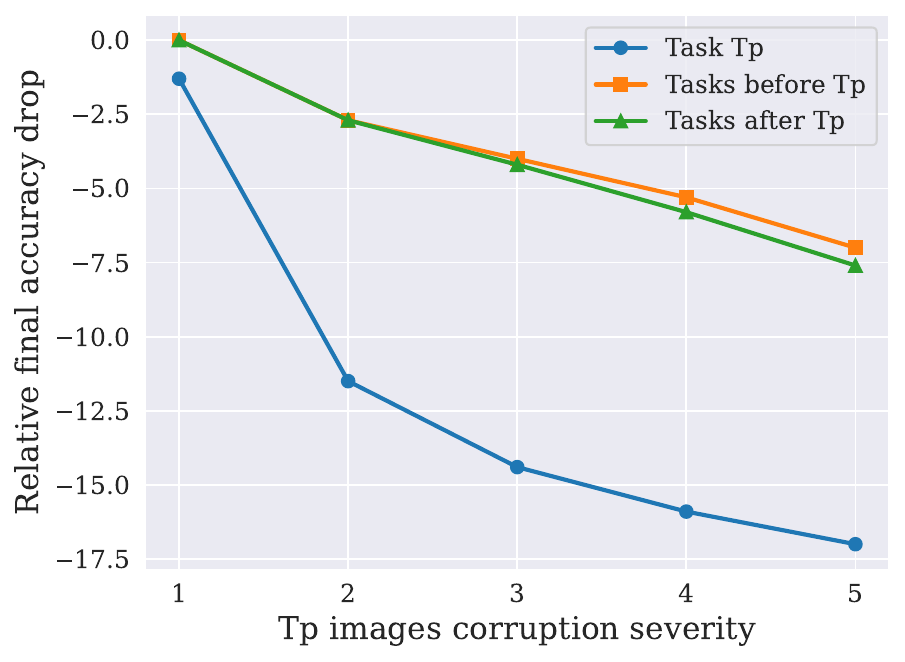}
        \caption{}
    \end{subfigure}
    \begin{subfigure}{0.47\textwidth}
        \includegraphics[width=\linewidth]{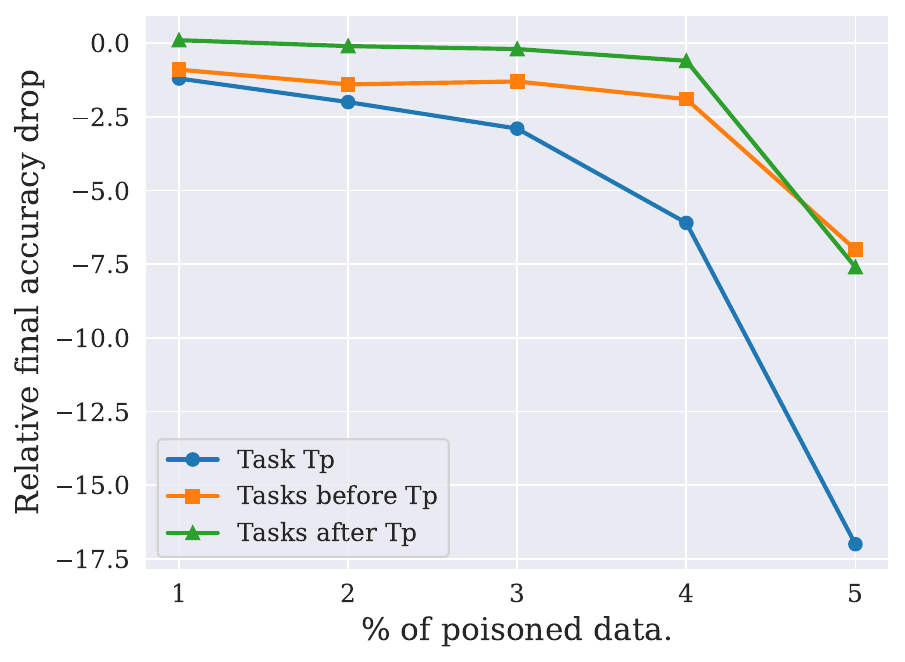}
        \caption{}
    \end{subfigure}
    \hfill
    
    \caption{\textbf{\Bait~attack effect for different corruption severities (a) and \% of poisoned data (b).} \textbf{(a)} Accuracy drop for past and future tasks correlates with image corruption severity. \textbf{(b)} Significant part of $T_p$ must be poisoned to create an effective attack. Note that we report \% of full training data for CIFAR100, 5\% corresponds to fully poisoning ($pp=100$) half of classes in $T_p$ ($pcp=50$), which corresponds to strongest possible \Bait~attack. Experimental setup: CIFAR100, LWF, Resnet32.}
    \label{fig:severity_results_pp_results}
\end{figure}

\minisection{4A) Adversary may use different corruption severities to adjust the attack strength.} Figure~\ref{fig:severity_results_pp_results}a) shows relative final accuracy drop for \Bait~attack with image corruptions of different severity (see Figure~\ref{fig:STP-corruptions-severities}). Corruptions with severity equal to 1 are negligible and can be seen as augmentation. Therefore only accuracy of $T_p$ is slightly affected. However, increasing severity correlates with increasing performance drop on all tasks. Thus, corruption severity can be seen as a kind of poisoning 'adversarial budget' for the attacker. Smaller severity means smaller performance drop, but also smaller chance of poison detection. 

\minisection{4B) Significant part of $T_p$ must be poisoned to create an effective attack.} 
Our additional experiments show that adversary must poison large fraction of samples in $T_p$ (Figure~\ref{fig:severity_results_pp_results}b)) to obtain a significant effect. Poisoning 4\% of data (i.e. 80\% of exemplars in half of the classes in $T_p$) is much less effective than poisoning 5\% (i.e. 100\% of exemplars in half of the classes in $T_p$). This may be seen as an attack limitation (because the attacker cannot control the strength of the attack by poisoning small part of the $T_p$ data). 
\newline

\subsection{Defense evaluation}
\begin{figure}
    \centering
    \begin{minipage}{0.47\textwidth}
        \centering
        \includegraphics[width=\linewidth]{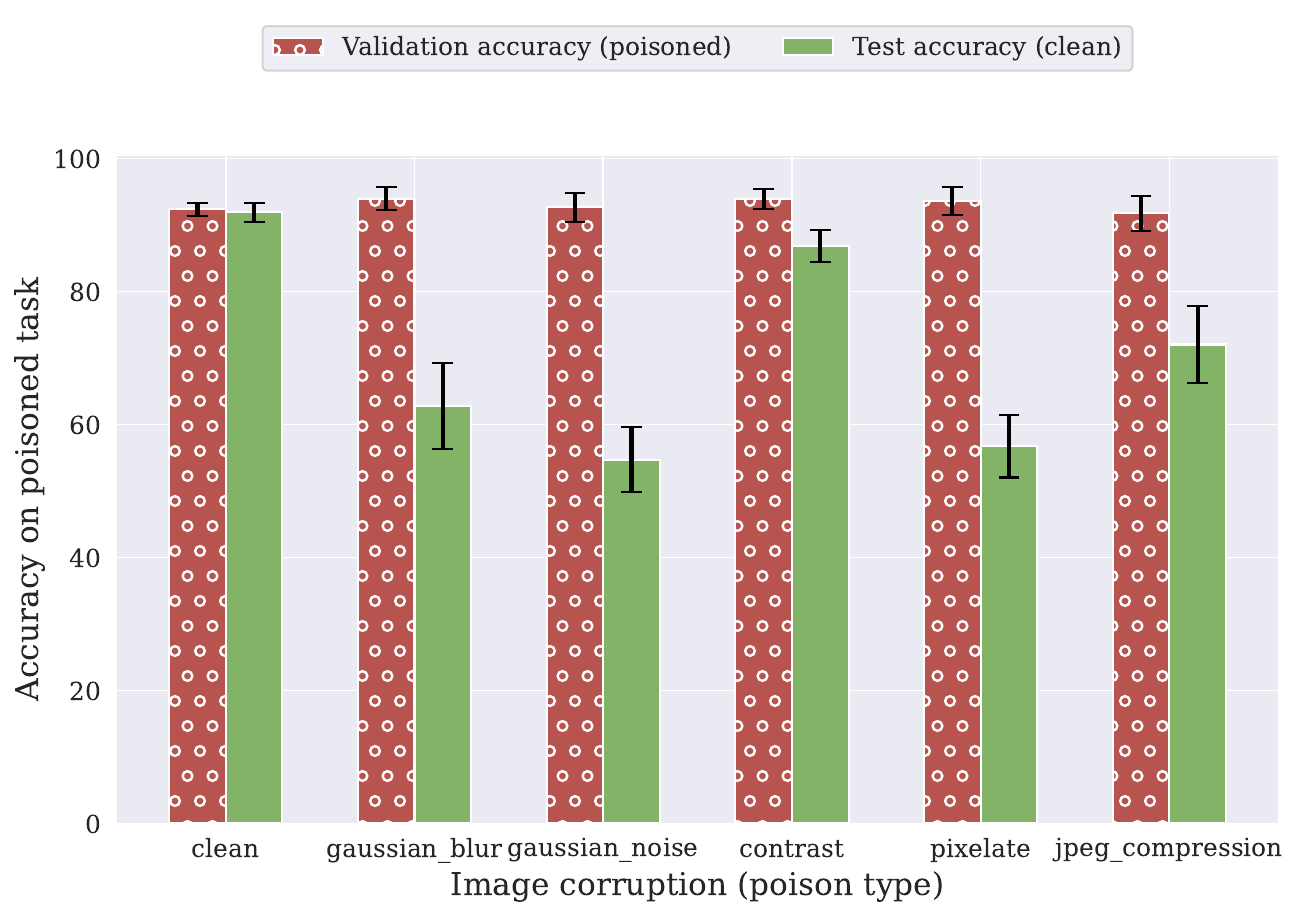}  \caption{\textbf{STP attacks obtain a reasonable validation accuracy, while real performance on test data is much worse.} While there is no difference between evaluation when $T_p$ is not poisoned ('clean' bars), there is significant drop for test data after attack. However, during CL training defender party has only access to (possibly) poisoned validation data, which cannot be simply used to detect the attack. Experimental setup: CIFAR100, LWF, Resnet32, $T_p = 4$.
  }
   \label{fig:val_test_evaluation_defense}
    \end{minipage}
    \hspace{0.5cm}
    \begin{minipage}{0.41\textwidth}
    \vspace{-0.5cm}
        \centering
        \includegraphics[width=\linewidth]{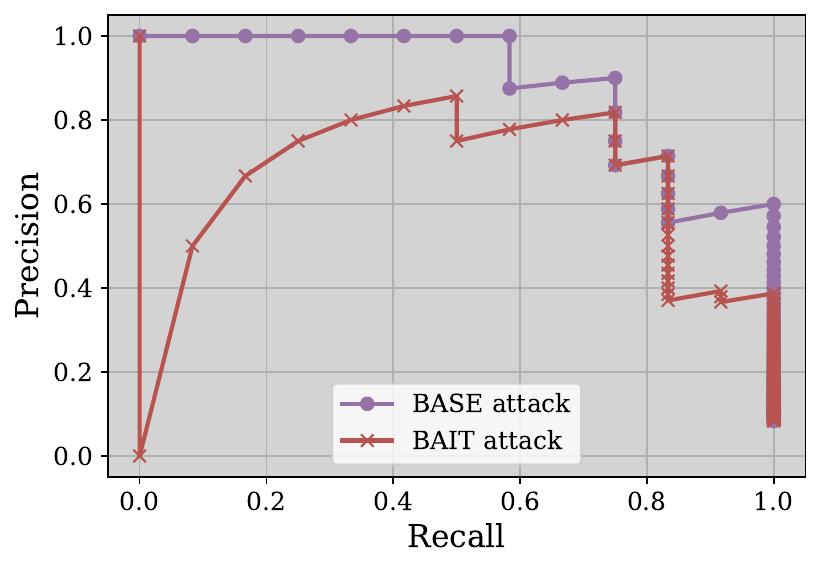}   
  \caption{\textbf{Precision-Recall curve for our defense method.} \Bait~attacks are harder to detect than \Base~attacks, due to poisoning less exemplars. Experimental setup: CIFAR100, LWF, Resnet32, $T_c = 2$, $T_p = 4$.}
   \label{fig:pr_curve_defense}
    \end{minipage}
\end{figure}

We explore defense against STP attacks via these questions:
\begin{enumerate}
\setcounter{enumi}{4}
\item Is the defending party able to actively detect poisoning during
learning task $T_p$ based on the validation accuracy?
\item How effective is the proposed poison task detection?
\end{enumerate}
\newpage
We answer them with the following observations:\newline
\minisection{5)~STP attacks obtain a reasonable validation accuracy
on task $T_p$.} An important characteristic of a successful data poisoning attack is it stealthiness, that ensures the attack is not easy to spot during training on the $T_p$ task. As the data poisoning attacks usually impede the training, the most basic defense mechanism is to monitor the accuracy of the classifier on a hold-out clean validation dataset representative to all classes. The defense against data poisoning attack is more challenging in CL. With no access to previous tasks’ data the defender party can only monitor the accuracy for the current task. Moreover, in realistic CL setup, the defender party does not have access to a clean validation dataset: it is obtained by splitting the data from a new task into train and validation sets. Then, if the task is poisoned with STP it affects also the validation set and the defender party cannot easily discover the attack only by monitoring current task
accuracy. Figure \ref{fig:val_test_evaluation_defense} supports these claims and show that the adversary obtains reasonable accuracy for poisoned task for all attacks. Note, that the defender party is not able to tell beforehand what should be the accuracy for an unknown task during CL training. While there is almost no difference between validation and test data in clean data case, the difference is significant for poisoned validation set and
clean test showing the data distribution change caused by image corruptions. Note that clean test data is not available for the defender party during training and here used only for evaluation purposes.

\minisection{6A)~Poison task detection using task vectors correctly identify most of the poisoned tasks.} Defender party is looking for a trade-off between the number of correctly detected attacks (True Positives) and false alarms when task is not poisoned (False Positives). This trade-off is connected with angle threshold selection (see Figure~\ref{fig:task-vectors-defense}). The bigger the angle, the less false alarms, but also less correct attack detections. These trade-off can be presented on Precision-Recall curve (see Figure~\ref{fig:pr_curve_defense}). While selecting the threshold and further defense evaluation is discussed in detail in the appendix~\ref{defense_additional_experiments}, we report results for the most straightforward approach taking the maximum clean angle calculated from one of the tasks from the beginning of CL sequence. For that threshold we obtain 92.3\% for \Base~and  86.2\% of total accuracy for \Bait~attacks (predicting correctly respectively 80\% and 60\% of attacks).\newline

\minisection{6B)~\Bait~attacks are harder to detect than \Base~attacks.} Results presented on Figure~\ref{fig:pr_curve_defense} confirm that \Base~attacks are easier to detect than \Bait. The main reason is that the latter poison only half of the classes in $T_p$, thus change in $T_p$ data distribution is more difficult to detect. 

\minisection{Limitations}
In this work we present the STP framework to investigate and mitigate data poisoning threats in CL. It is beneficial to start building defensive mechanisms from simplest possible setup, where adversary can poison only one task in the sequence, but it comes with a cost. STP is useful when investigating memory-free CL methods, but less with those using exemplars. In STP attack only one task can be poisoned, thus CL methods with memory have (a non-realistic) guarantee during $T_p$ task that all stored previous samples are clean, which counter STP attack by making it more similar to the case of joint training (see results in appendix~\ref{attack_additional_experiments}). 
\section{Conclusion}
In this work, we propose Single-Task Poison (STP) to investigate data poisoning attacks in Continual Learning. In contrast to previously proposed poisoning settings, in STP adversaries do not have knowledge and access to a model, previous and future tasks data, having access only to a single task in the data stream. Our study demonstrates that even within these stringent conditions adversaries can compromise model performance by poisoning data with basic image corruptions. We show that STP attacks strongly disrupt the training: decrease the performance for the past tasks and impede training on future tasks. Finally, we propose a high-level defense framework for CL and evaluate a poison task detection method based on task vectors. Our research reveals CL vulnerabilities to data poisoning attacks and supports further investigation of such threats to address security challenges present in continual learning with adequate defense methods.

\paragraph{Acknowledgements.} 
This research was supported by Warsaw University of Technology (Poland) within the Excellence Initiative Research University (IDUB) programme. We additionally acknowledge projects PID2022-143257NB-I00, financed by MCIN/AEI/10.13039/501100011033 and FSE+, funding by the European Union ELLIOT project, and the Generalitat de Catalunya CERCA Program. Bartłomiej Twardowski acknowledges the grant RYC2021-032765-I and National Centre of Science (NCN, Poland) Grant No. 2023/51/D/ST6/02846. Stanisław Pawlak acknowledges National Centre of Science (NCN, Poland) Grant No. 2023/51/D/ST6/01609.

\bibliography{collas2025_conference}
\bibliographystyle{collas2025_conference}
\newpage
\section{Appendix}
\appendix
\section{Additional experiments}
\subsection{Extended investigation of STP Attacks}\label{attack_additional_experiments}
We state the following questions to further investigate STP attacks:
\begin{enumerate}
\setcounter{enumi}{6}
\item How plasticity-stability regularization affect the outcome of the attack? Which task in the sequence is the easiest to poison?
\item How does connection between different severities of the corruption and different percentage of poisoned exemplars ($pp$) affect STP attack strength?
\item Are STP attacks successful for a stream containing more complex images than those in CIFAR10/100?
\item Are STP attacks successful when using memory-based CL methods?
\end{enumerate}

We answer them with the following observations:\\

\begin{figure}[h!]
  \centering
  \includegraphics[width=0.5\textwidth]{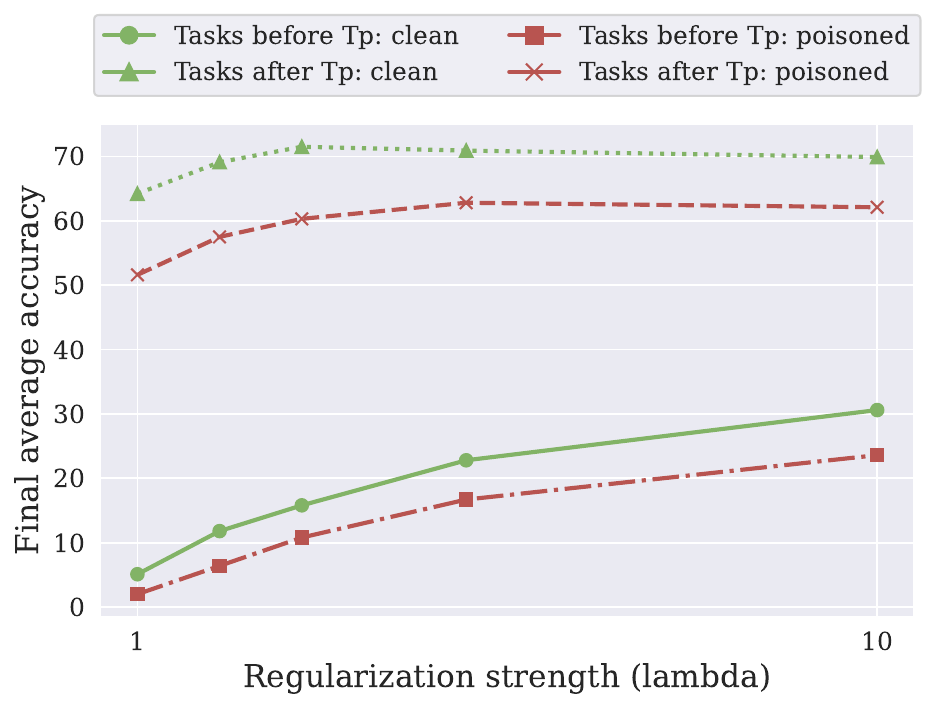}   
  \caption{\textbf{Regularization strength matters in STP attack.}The stronger regularization, the smaller is the impact of poisoning for future tasks performance, while bigger for past tasks.
  }
   \label{fig:lambdas}
\end{figure}

\minisection{7A) Regularization strength matters in STP attacks.} Increasing the regularization in CL improves model performance for past tasks. Thus, it should also affect the outcome of the STP attack. Figure~\ref{fig:lambdas} shows that increasing regularization increase also the poisoning impact on past tasks, while undermines the impact of poisoning for future tasks performance. The relative performance drop for past tasks increases from -3.1 at $\lambda=1$ to -7.1 at $\lambda=10$. Conversely, for future tasks, the relative performance drop decreases from -12.6 at $\lambda=1$ to -7.4 at $\lambda=10$. This result is consistent with the fact, that the ability to poison the model is connected with the plasticity of the model. For indiscriminate attacks, the lower the plasticity, the lower is the data poisoning risk. In an extreme case, when the plasticity is equal to zero (model is not updated, the weights are fixed), poisoning the data has no effect.

\begin{figure}[bt!]
    \centering
    
    \begin{subfigure}{0.3\textwidth}
        \includegraphics[width=\linewidth]{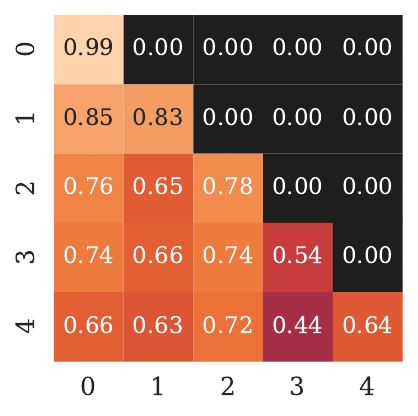}
    \end{subfigure}
    \hfill
    \begin{subfigure}{0.3\textwidth}
        \includegraphics[width=\linewidth]{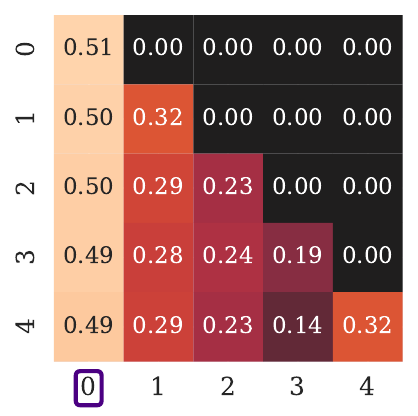}
    \end{subfigure}
    \hfill
    \begin{subfigure}{0.3\textwidth}
        \includegraphics[width=\linewidth]{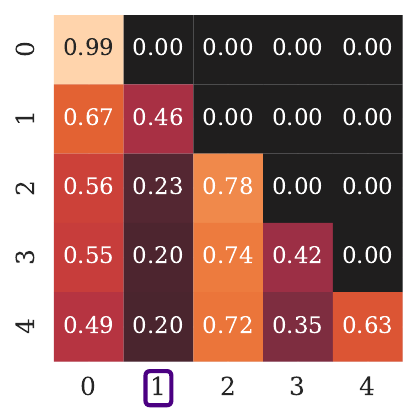}
    \end{subfigure}
        \hfill
    \begin{subfigure}{0.3\textwidth}
        \includegraphics[width=\linewidth]{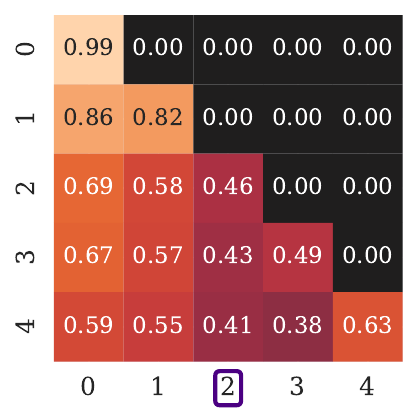}
    \end{subfigure}
    \hfill
    \begin{subfigure}{0.3\textwidth}
        \includegraphics[width=\linewidth]{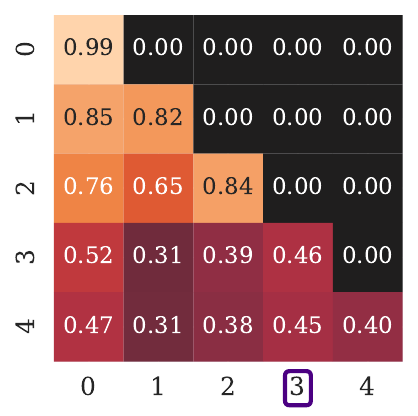}
    \end{subfigure}
    \hfill
    \begin{subfigure}{0.3\textwidth}
        \includegraphics[width=\linewidth]{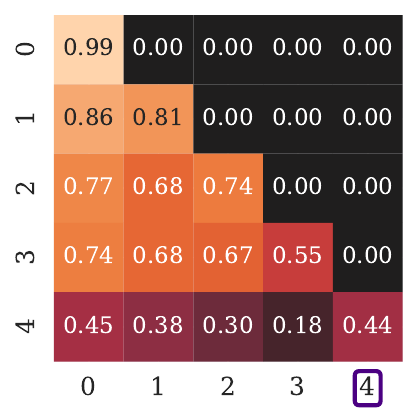}
    \end{subfigure}
    
    \caption{\textbf{Poisoning different tasks on CIFAR10: \Bait~attack with gaussian blur corruption.} Top left image is LwF class incremental accuracy on clean data. Other plots have poisoned $T_{p}$ number marked with purple frame.}

    \label{fig:lwf_cifar10_poison_tasks_discriminatory}
\end{figure}

\minisection{7B) Poisoning first task in a sequence has the most severe consequences.} Figure~\ref{fig:lwf_cifar10_poison_tasks_discriminatory} shows average accuracies for 5 tasks in CL split-CIFAR10 sequence with \Bait~attacks on $T_p$, where $p \in <1,5>$. When changing the poisoned task index $p$, we can observe that $T_p$ is the most affected task, consistently with our previous claim in observation \textbf{2A}. Moreover, poisoning the first task in the sequence is far the most influential for further training: the model trained on poisoned data from the start struggles to learn significant patterns later on. While interesting, the significance of this fact is not crucial: the defender party has much bigger control on the first task or model initialization than any other task in the sequence, so additional defensive mechanism can be used to eliminate this threat.

\begin{figure*}[bt]
    \centering
    
    \begin{subfigure}{0.3\textwidth}
        \includegraphics[width=\linewidth]{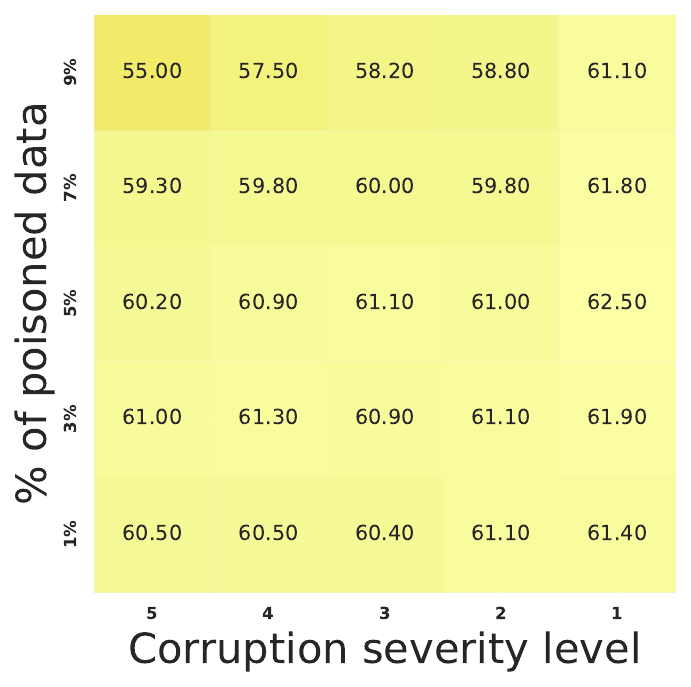}
        \caption{}
    \end{subfigure}
    \begin{subfigure}{0.3\textwidth}
        \includegraphics[width=\linewidth]{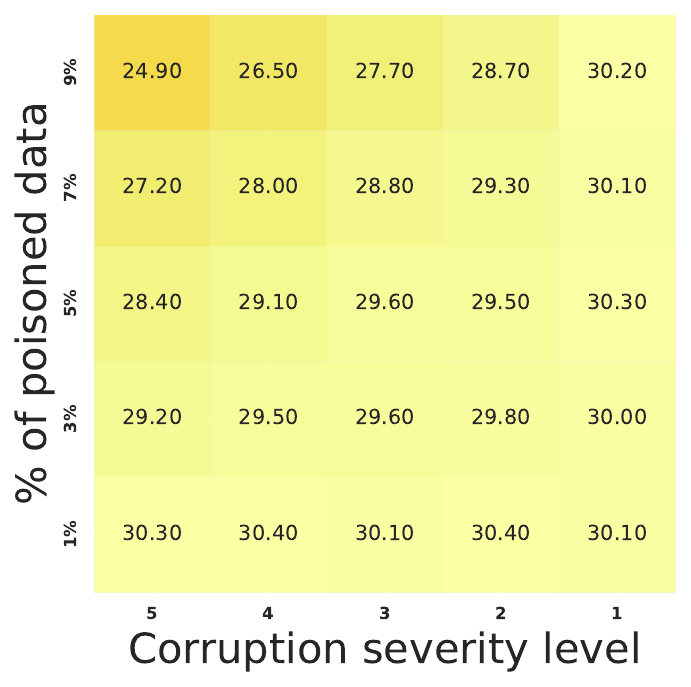}
        \caption{}
    \end{subfigure}
     \begin{subfigure}{0.3\textwidth}
        \includegraphics[width=\linewidth]{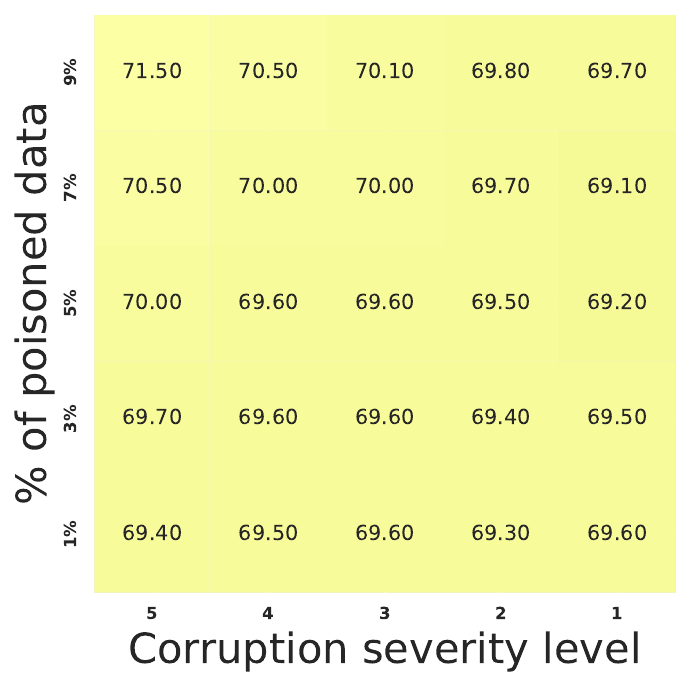}
        \caption{}
    \end{subfigure}
    \hfill
    \begin{subfigure}{0.3\textwidth}
        \includegraphics[width=\linewidth]{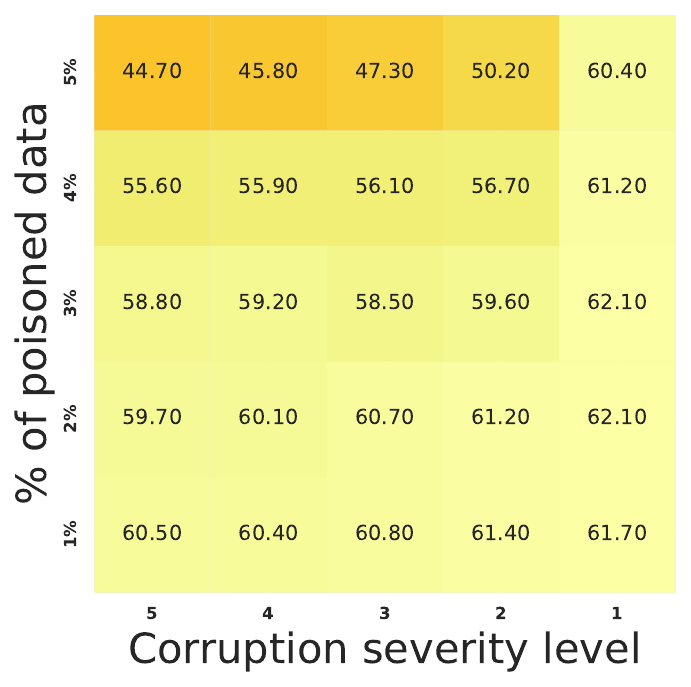}
        \caption{}
    \end{subfigure}
    \begin{subfigure}{0.3\textwidth}
        \includegraphics[width=\linewidth]{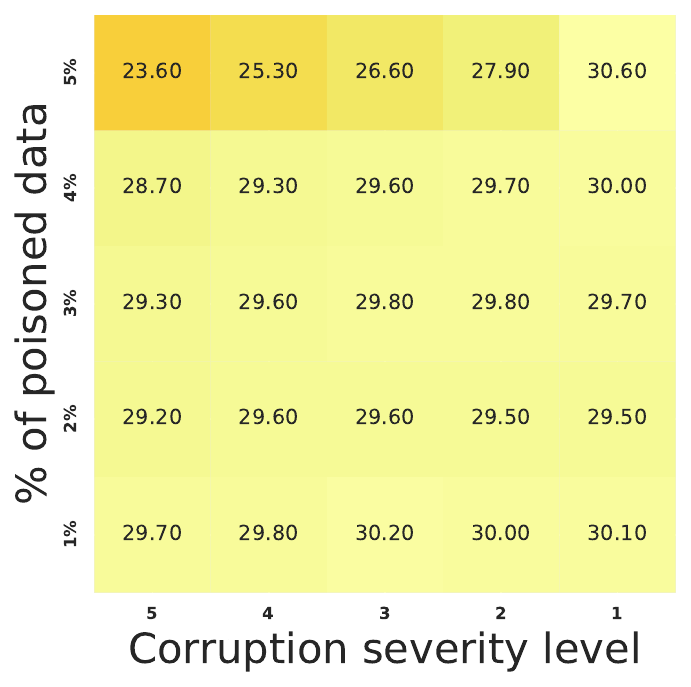}
        \caption{}
    \end{subfigure}
     \begin{subfigure}{0.3\textwidth}
        \includegraphics[width=\linewidth]{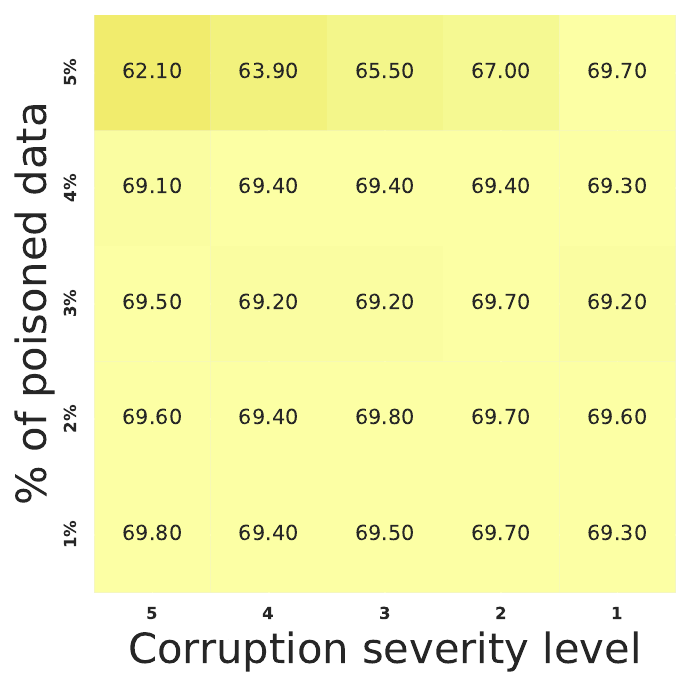}
        \caption{}
    \end{subfigure}
    \hfill
    
    \caption{\textbf{The impact of corruption severity and poisoned percentage of samples on STP attack performance.} Poisoning a vast percentage of class exemplars is more important than corruption severity. We report the average accuracy for: $T_p$, $BEFORE~T_p$ and $AFTER~T_p$ for \Base~attack (a-c) and \Bait~attack (d-f). Note that the \% on y axis is calculated as \% of samples poisoned in the whole dataset and 10\% means that all samples in $T_p$ are poisoned.}
    \label{fig:corruption_vs_pp}
\end{figure*}

\minisection{8) Poisoning a vast percentage of class exemplars is more important than corruption severity.}
Figure~\ref{fig:corruption_vs_pp} is an extension of Figure~\ref{fig:severity_results_pp_results} presented in the main body of this work. It shows (in accordance to observations stated in \textbf{4)}) that two conditions are necessary for a successful STP attack. Firstly, the corruption of the image itself cannot be negligible. Very low severity does not disrupt the training (\textbf{4A}). Secondly, the vast percentage of $T_p$ must be poisoned (\textbf{4B}). From those two conditions, the latter has greater significance. When poisoning $>= 90\%$ of class exemplars in $T_p$ performance decrease even with smaller severities of corruption. On the other hand, when less exemplars are poisoned the accuracy drop is negligible even for the strongest severity considered.


\definecolor{mygray}{gray}{0.5}
\addtolength{\tabcolsep}{-3.5pt} 
\begin{table*}[t]
\begin{center}
\caption{\textbf{Average accuracy when poisoning a single $T_p$ task in CL sequence.} \Clean~row shows results of runs without poisoning $T_p$ task, while black (\Base, \Bait, \Multibase, \Multibait) rows show results for different attacks. Difference between \Clean~and poisoned runs is shown in parentheses. We report accuracies after the poisoned task (ACC AT POISONING TIME). Additionally, we report accuracies at the end of the full CL sequence (FINAL ACC) to show the impact of poisoning for further training.}
\label{tab:lwf_results_tiny}
\begin{sc}
\begin{tabular}{lccc|ll|llll}
\toprule
Attack & Method & Dataset & p & \multicolumn{2}{c}{Acc at poisoning time} & \multicolumn{4}{c}{Final Acc} \\
\cmidrule(r){5-6} \cmidrule(r){7-10}
&&&&  $T_p$ &  Before $T_p$ &  $T_p$ &  Before $T_p$ &  After $T_p$ & Total \\
\midrule
\Clean & LWF & TINY & 4 & 75.8 & 39.8 & 47.3 & 23.0 & 65.9 & 41.3\\
\cdashlinelr{1-10}
\Base & LWF & TINY & 4 & 43.5 \tiny{(-32.3)} & 21.2 \tiny{(-18.6)} & 23.3 \tiny{(-24.0)} & 12.5 \tiny{(-10.5)} & 58.6 \tiny{(-7.3)} & 29.7 \tiny{(-11.6)}\\
\Bait & LWF & TINY & 4 & 45.2 \tiny{(-30.6)} & 32.5 \tiny{(-7.3)} & 33.9 \tiny{(-13.4)} & 17.5\tiny{(-5.5)} & 62.0 \tiny{(-3.9)} & 35.1 \tiny{(-6.2)}\\
\Multibase & LWF & TINY & 4 & 64.2 {\tiny (-11.6)} & 39.1 {\tiny (-0.7)} & 41.3 {\tiny (-6.0)} & 21.9 {\tiny (-1.1)} & 64.6 {\tiny (-1.3)} & 39.4 {\tiny (-1.9)} \\
\Multibait & LWF & TINY & 4 &  65.8 {\tiny (-10.0)} & 39.8 {\tiny (0.0)} & 38.4 {\tiny (-8.9)} & 21.6 {\tiny (-1.4)} & 59.2 {\tiny (-6.7)} & 36.9 {\tiny (-4.4)}\\
\bottomrule
\end{tabular}
\end{sc}
\end{center}

\end{table*}

\begin{table}[ht!]
\begin{center}
\caption{\textbf{Poison task detection for \Base~attacks.} P90 achieves the highest overall accuracy and the top F1 score. THRESHOLD -- statistic used to calculate $\alpha$, ACC -- total detection accuracy (on clean and poisoned tasks), CLEAN -- accuracy on clean tasks, ATTACK -- accuracy on poisoned tasks, F1 -- F1 score.}
\label{tab:eval_thres}
\begin{sc}
\begin{tabular}{lllll}
\toprule
Threshold &  ACC & Clean & Attack & F1 \\
\midrule
MAX+IQR &  \textbf{93.8} & \textbf{100.0} & 80.0 & 0.86 \\
MAX & 92.3 & 97.8 & 80.0 & 0.86 \\
P90 &  \textbf{93.8} & 97.8 & 85.0 & \textbf{0.89} \\
MAX-IQR & 92.3 & 91.1 & \textbf{95.0} & 0.88 \\
P75  & 84.6 & 80.0 & \textbf{95.0} & 0.79 \\
\bottomrule
\end{tabular}
\end{sc}
\end{center}

\end{table}

\begin{table}[ht!]
\begin{center}
\caption{\textbf{Poison task detection for \Bait~attacks.} P90 achieves the highest overall accuracy and the top F1 score. THRESHOLD -- statistic used to calculate $\alpha$, ACC -- total detection accuracy (on clean and poisoned tasks), CLEAN -- accuracy on clean tasks, ATTACK -- accuracy on poisoned tasks, F1 -- F1 score.}
\label{tab:eval_thres_bait}
\begin{sc}
\begin{tabular}{lllll}
\toprule
Threshold & ACC & Clean & Attack & F1 \\
\midrule
MAX+IQR & 75.4 & \textbf{100.0} & 20.0 & 0.10 \\
MAX & 86.2 & 97.8 & 60.0 & 0.73 \\
P90 & \textbf{89.2} & 97.8 & 70.0 & \textbf{0.80} \\
MAX-IQR & 84.6 & 91.1 & 70.0 & 0.74 \\
P75 & 80.0 & 80.0 & \textbf{80.0} & 0.71 \\
\bottomrule
\end{tabular}
\end{sc}
\end{center}

\end{table}

\begin{table*}[t]
\begin{center}
\caption{\textbf{Average accuracy when poisoning a single $T_p$ task in CL sequence with memory buffer of 2000 exemplars.} \Clean~rows show results of runs without poisoning $T_p$ task, while black (\Base, \Bait) rows show results for different attacks. Difference between \Clean~and poisoned runs is shown in parentheses. We report accuracies after the poisoned task (ACC AT POISONING TIME). Additionally, we report accuracies at the end of the full CL sequence (FINAL ACC) to show the impact of poisoning for further training.}
\label{tab:buffer_results}
\begin{sc}
\begin{tabular}{lccc|ll|llll}
\toprule
Attack & Method & Dataset & p & \multicolumn{2}{c}{Acc at poisoning time} & \multicolumn{4}{c}{Final Acc} \\
\cmidrule(r){5-6} \cmidrule(r){7-10}
&&&&  $T_p$ &  Before $T_p$ &  $T_p$ &  Before $T_p$ &  After $T_p$ & Total \\
\midrule
\Clean & FINETUNING & CIFAR10 & 3 & 97.6 & 84.5 & 74.8 & 73.9 & 86.4 & 79.1 \\
\cdashlinelr{1-10}
\Base &  FINETUNING & CIFAR10 & 3 & 0.1 \tiny{(-97.5)} & 93 \tiny{(8.5)} & 0.0 \tiny{(-74.8)} & 74.7 \tiny{(0.8)} & 86.9 \tiny{(0.5)} & 64.6 \tiny{(-14.5)} \\
\Bait &  FINETUNING & CIFAR10 & 3 & 49.0 \tiny{(-48.6)} & 87.2 \tiny{(2.7)} & 35.2 \tiny{(-39.6)} & 73.8\tiny{(-0.1)} & 86.9 \tiny{(0.5)} & 71.3 \tiny{(-7.8)}\\
\midrule
\Clean & FINETUNING & CIFAR100 & 4 & 90.4 & 44.9 &42.9 & 37.5 & 64.3 & 47.4\\
\cdashlinelr{1-10}
\Base &  FINETUNING & CIFAR100 & 4 & 0.1 \tiny{(-90.3)} & 56.9 \tiny{(12.0)} & 0.0 \tiny{(-42.9)} & 38.1 \tiny{(0.6)} & 64.7 \tiny{(0.4)} & 40.6 \tiny{(-6.8)}\\
\Bait &  FINETUNING & CIFAR100 & 4 & 45.8 \tiny{(-44.6)} & 50.2 \tiny{(5.3)} & 20.5 \tiny{(-22.4)} & 37.8\tiny{(0.3)} & 64.2 \tiny{(-0.1)} & 43.7 \tiny{(-3.7)}\\
\midrule
\midrule
\\
Clean & GDUMB & CIFAR10 & 3 & 76.3 & 78.7 & 68.0 & 66.5 & 66.7 & 66.9\\
\cdashlinelr{1-10}
\Base &  GDUMB & CIFAR10 & 3 & 0.7 \tiny{(-75.6)} & 83.9 \tiny{(5.3)} & 0.9 \tiny{(-67.1)} & 71.4 \tiny{(4.9)} & 69.1 \tiny{(2.4)} & 56.4 \tiny{(-6.8)}\\
\Bait &  GDUMB & CIFAR10 & 3 & 39.0 \tiny{(-37.3)} & 80.7 \tiny{(2.0)} & 34.4 \tiny{(-33.6)} & 68.8\tiny{(2.3)} & 67.1 \tiny{(0.4)} & 61.3\tiny{(-5.6)}\\
\midrule
\Clean & GDUMB & CIFAR100 & 4 & 23.8 & 26.0 & 21.5 & 21.8 & 17.8 & 20.4\\
\cdashlinelr{1-10}
\Base &  GDUMB & CIFAR100 & 4 & 0.2 \tiny{(-23.6)} & 28.4 \tiny{(2.4)} & 0.5\tiny{(-21.0)} & 23.0 \tiny{(1.2)} & 19.5 \tiny{(1.7)} & 18.1 \tiny{(-2.3)}\\
\Bait &  GDUMB & CIFAR100 & 4 & 13.9 \tiny{(-9.9)} & 27.3 \tiny{(1.3)} & 12.1 \tiny{(-9.4)} & 22.9\tiny{(1.1)} & 17.9 \tiny{(0.1)} & 19.5 \tiny{(-0.9)}\\
\bottomrule
\end{tabular}
\end{sc}
\end{center}

\end{table*}

\begin{table*}[bt]
\begin{center}
\caption{\textbf{Average accuracy when poisoning a single $T_p$ task in CL sequence.} \Clean~rows show results of runs without poisoning $T_p$ task, while black (\Base, \Bait, \Multibase, \Multibait) rows show results for different attacks. Difference between \Clean~and poisoned runs is shown in parentheses. We report accuracies after the poisoned task (ACC AT POISONING TIME). Additionally, we report accuracies at the end of the full CL sequence (FINAL ACC) to show the impact of poisoning for further training.}
\label{tab:lwm_results}
\begin{sc}
\begin{tabular}{lccc|ll|llll}
\toprule
Attack & Method & Dataset & p & \multicolumn{2}{c}{Acc at poisoning time} & \multicolumn{4}{c}{Final Acc} \\
\cmidrule(r){5-6} \cmidrule(r){7-10}
&&&&  $T_p$ &  Before $T_p$ &  $T_p$ &  Before $T_p$ &  After $T_p$ & Total \\
\midrule
\Clean & LWM & CIFAR10 & 3 & 67.8 & 71.0 & 51.4 & 62.6 & 62.8 & 60.4\\
\cdashlinelr{1-10}
\Base & LWM & CIFAR10 & 3 & 55.6 {\tiny (-12.2)} & 69.8 {\tiny (-1.2)} & 40.0 {\tiny (-11.4)} & 58.2 {\tiny (-4.4)} & 64.5 {\tiny (1.7)} & 57.1 {\tiny (-3.3)}\\
\Bait & LWM & CIFAR10 & 3 & 45.5 {\tiny (-22.3)} & 50.5 {\tiny (-20.5)} & 41.8 {\tiny (-9.6)} & 45.8 {\tiny (-16.8)} & 50.2 {\tiny (-12.6)} & 46.8 {\tiny (-13.6)}\\
\midrule
\midrule
\Clean & LWM & CIFAR100 & 4 & 79.0 & 51.1 & 62.2 & 31.1 & 70.1 & 49.3\\
\cdashlinelr{1-10}
\Base & LWM & CIFAR100 & 4 & 64.4 {\tiny (-14.6)} & 37.7 {\tiny (-13.4)} & 46.7 {\tiny (-15.5)} & 18.1 {\tiny (-13.0)} & 69.4 {\tiny (-0.7)} & 40.0 {\tiny (-9.3)}\\
\Bait & LWM & CIFAR100 & 4 & 52.3 {\tiny (-26.7)} & 40.7 {\tiny (-10.4)} & 45.1 {\tiny (-17.1)} & 23.6 {\tiny (-7.5)} & 62.1 {\tiny (-8.0)} & 40.0 {\tiny (-9.3)}\\
\Multibase & LWM & CIFAR100 & 4 & 36.6 {\tiny (-42.4)} & 45.9 {\tiny (-5.2)} & 28.6 {\tiny (-33.6)} & 26.8 {\tiny (-4.3)} & 65.2 {\tiny (-4.9)} & 39.9 {\tiny (-9.4)}\\
\Multibait & LWM & CIFAR100 & 4 & 54.5 {\tiny (-24.5)} & 47.5 {\tiny (-3.6)} & 47.2 {\tiny (-15.0)} & 28.4 {\tiny (-2.7)} & 65.7 {\tiny (-4.4)} & 44.0 {\tiny (-5.3)}\\
\bottomrule
\end{tabular}
\end{sc}
\end{center}
\end{table*}

\minisection{9A) STP attacks are successful for streams with more complex images.} To confirm that STP attacks pose a threat also for more complex CL benchmarks, we conduct experiments on Split Tiny Imagenet. Table~\ref{tab:lwf_results_tiny} shows similar results to those presented in main part of this work (see Table~\ref{tab:main_results}): in terms of FINAL ACC, the most affected is accuracy on poisoned task ($T_p$), but there exists also significant drop for $BEFORE~T_p$ and $AFTER~T_p$ results. The results further reinforce the evidence supporting the effectiveness of STP attacks and validate their impact on more complex datasets beyond CIFAR-10/100.

\minisection{9B) \Base~attacks are stronger with increasing number of classes in $T_p$ and increasing classification complexity.} 
Table~\ref{tab:lwf_results_tiny} shows strong performance drop for \Base~ attack. We attribute it to two following factors: 1) the number of classes in $T_p$, 2) total number of classes (i.e. classification task complexity). Regarding the TOTAL FINAL ACC, the decreases become more pronounced as the number of classes and dataset complexity increase: -2.7 (with $T_p$ having two classes for CIFAR-10), -9.7 (with $T_p$ having 10 classes for CIFAR-100), and -11.6 (with $T_p$ having 20 classes for Tiny ImageNet).

\minisection{10) Performance on past and future tasks is not affected for STP attacks on memory-based CL methods.}
In the Limitations section of this work, we note that the STP framework is effective for investigating memory-free CL methods but less applicable to those utilizing exemplars. The STP framework restricts the adversary to poisoning only a single task. When only one task is poisoned, CL methods with memory maintain an unrealistic guarantee that all previous samples stored in the buffer are clean during the $T_p$ task (since $T_p$ is the only poisoned task). Introducing an additional buffer of clean samples undermines the concept of 'learning the task in isolation' and mitigates the STP attack, thereby creating a scenario more akin to joint training. Results for simple finetuning with buffer and GDUMB~(\cite{prabhu2020greedy}) presented in Table~\ref{tab:buffer_results} confirm that (similarly to JOINT training) attack decrease only performance on $T_p$, while there are no accuracy drops for $BEFORE~T_p$ and $AFTER~T_p$ results (numbers in parenthesis are positive or close to zero).

\minisection{Additional results for LwM.}
Table~\ref{tab:lwm_results} presents STP attacks evaluation for additional data regualrization method -- LwM~(\cite{dhar2019learning}). These results are consistent with observations described in the main part of this work.
\begin{figure}[bt]
  \centering
  \includegraphics[width=0.48\textwidth]{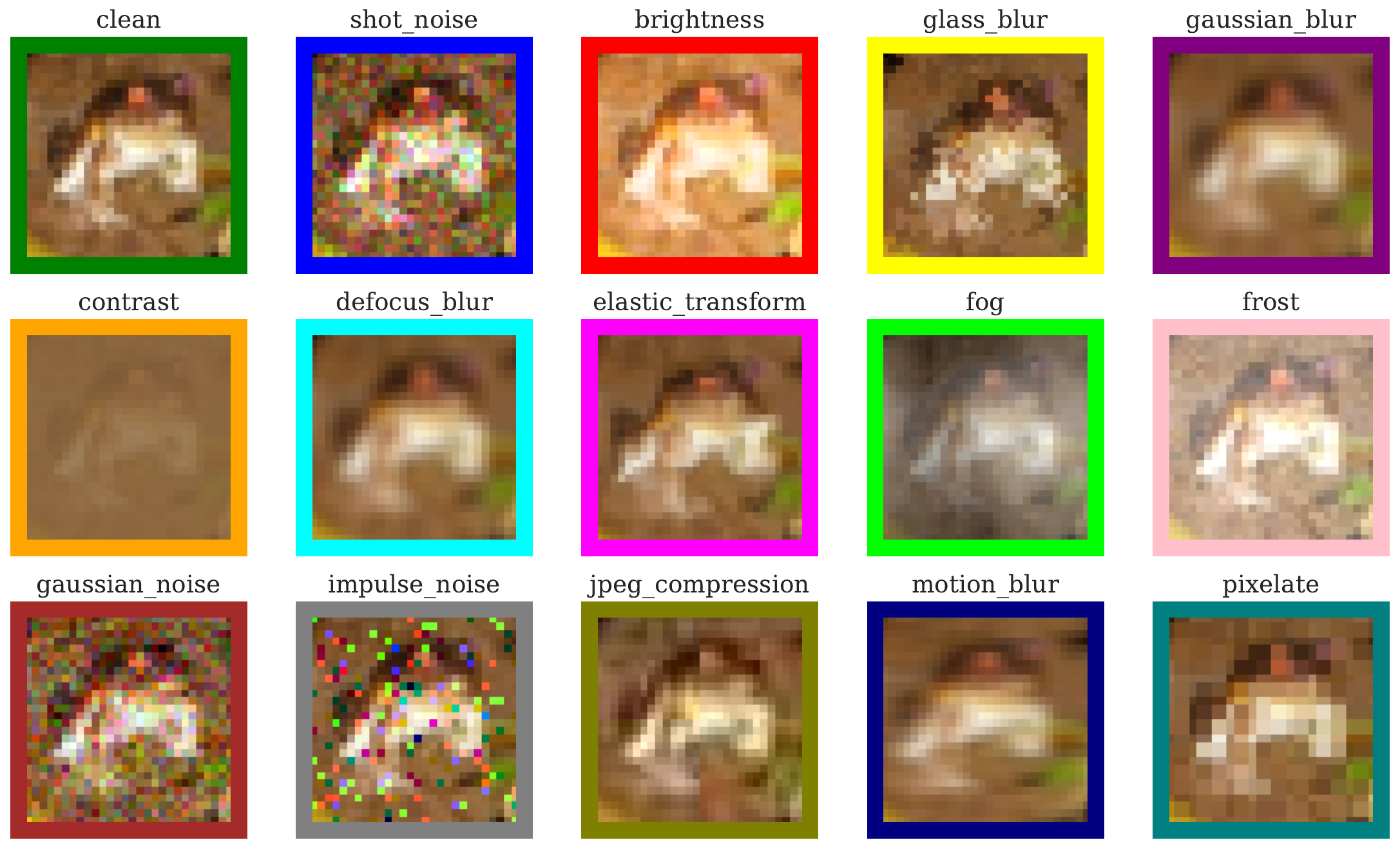}   
  \caption{\textbf{Image corruptions used to create poisoned training samples.} We use CIFAR-C set of image corruptions to transform images of the poisoned task. We show clean image in the upper right corner for comparison. We present corruptions with maximum severity (equal to 5).
  }
   \label{fig:STP-corruptions}
\end{figure}
\subsection{Extended investigation of defense against STP attacks.}\label{defense_additional_experiments}

The threshold value that determines whether a task is poisoned plays a crucial role in our method. We establish this threshold by using a clean task at the beginning of the training sequence. In order to obtain a threshold value we follow two simple steps: 1) We calculate the angle $\alpha'$ (see Figure~\ref{fig:task-vectors-defense}) on single clean task multiple times (using different random seeds for initialization), 2) We use a statistic measure to consolidate multiple possible threshold values into one $\alpha$. Aggregated threshold value is then applied to all subsequent tasks in the stream to assess whether they are poisoned. The choice of an appropriate threshold involves balancing the trade-off between correctly identifying attacks (True Positives) and minimizing false alarms when a task is not poisoned (False Positives). In the following section, we evaluate five different potential thresholds.

\minisection{Evaluation of different threshold values} 
We evaluate the following statistics to aggregate multiple $\alpha'$ values into single threshold $\alpha$:
\begin{itemize}
    \item MAX+IQR -- We calculate Interquartile Range for $\alpha'$ values and add it to maximum $alpha'$ value. This threshold is optimized for the lowest number of False Positives (false alarms).
    \item MAX -- maximum $alpha'$ value. 
    \item P90 -- We calculate 90th percentile of $alpha'$ values. 
    \item MAX-IQR -- We calculate Interquartile Range for $\alpha'$ values and subtract it from maximum $alpha'$ value. 
    \item P75 -- We calculate 75th percentile (Q3) of $alpha'$ values. This threshold is optimized for the highest number of True Positives (detected attacks).
\end{itemize}

Tables~\ref{tab:eval_thres} and~\ref{tab:eval_thres_bait} present the evaluation of the five proposed thresholds. The extreme options optimize performance for either clean data (MAX+IQR) or poisoned data (P75). While the choice of threshold ultimately depends on the specific needs of the defending party, metrics such as the F1 score can be used to achieve an appropriate balance between Precision and Recall. In our experiments, which considered five possible thresholds, P90 yielded the highest overall accuracy and the best F1 score, making it the most effective threshold among those evaluated.
\newpage
\subsection{Additional analysis and t-SNE visualizations of STP.}

\begin{figure}[h]
    \centering
     \begin{subfigure}{0.45\textwidth}
        \centering
           \includegraphics[width=\textwidth]{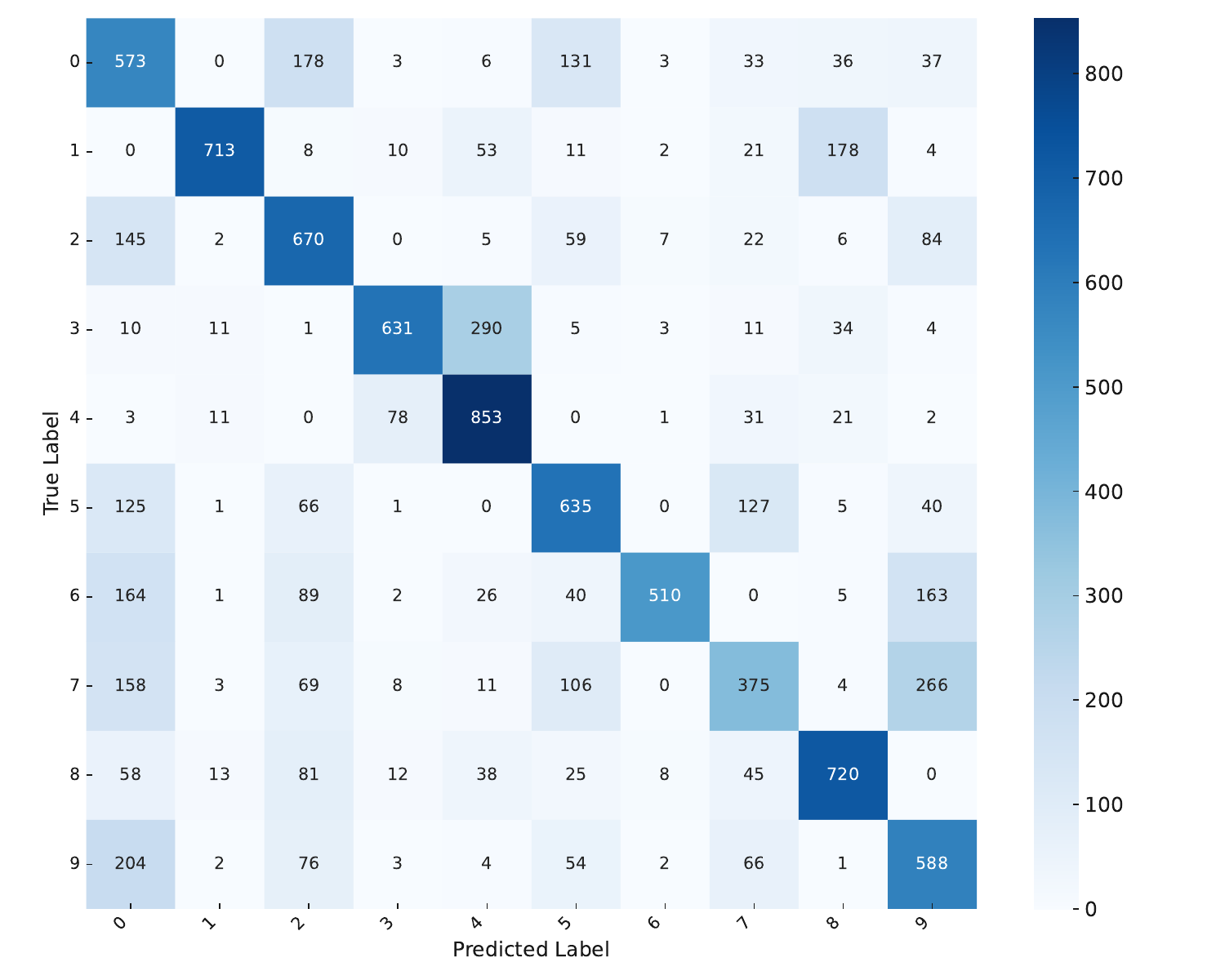}

          \caption{\Clean}
    \end{subfigure}
     \begin{subfigure}{0.45\textwidth}
        \centering
           \includegraphics[width=\textwidth]{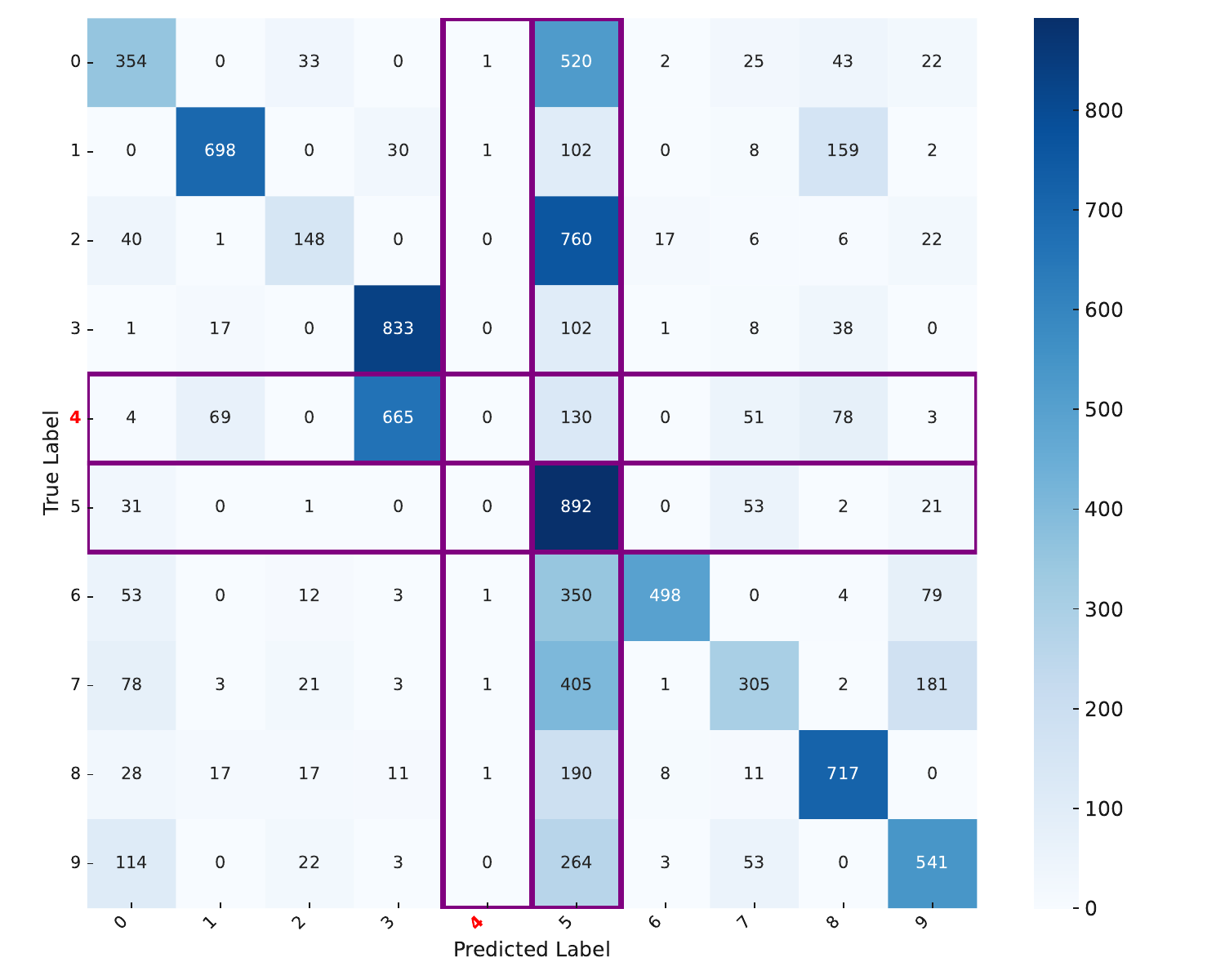}

           \caption{\Bait~attack}
    \end{subfigure}

    \caption{\textbf{Confusion matrices for CIFAR10 classes at the end of the CL training.} Clean run (left) consequently has high values on the diagonal, with low number of prediction errors. \Bait~attack (right) on the 2nd task results in almost no predictions for class with corrupted images (class 4, in red), while class 5 has very high number of wrong predictions.}
    \label{fig:conf_matrices}
    
\end{figure}
\minisection{\Bait~affects classes in the poisoned task unequally.}
As shown in Figure~\ref{fig:conf_matrices}, confusion matrices at the end of training reveal class-specific effects of the \Bait~attack. In Task 3, where two classes are involved, the class with label 4 (corrupted images) is almost never predicted -- its diagonal entry is effectively zero. Conversely, class 5 (uncorrupted) is over-predicted, reducing overall accuracy for other classes. This indicates that the attack's effect depends strongly on which class receives the corruption.

\begin{figure}[h!]
    \centering
     \begin{subfigure}{0.32\textwidth}
        \centering
           \includegraphics[width=\textwidth]{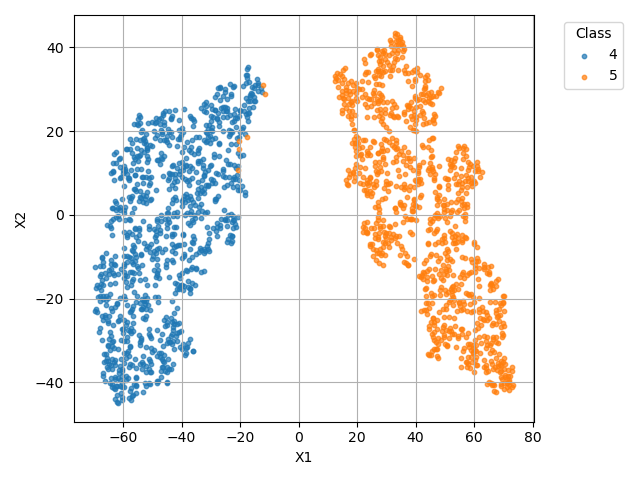}

          \caption{After 2nd task}
    \end{subfigure}
     \begin{subfigure}{0.32\textwidth}
        \centering
           \includegraphics[width=\textwidth]{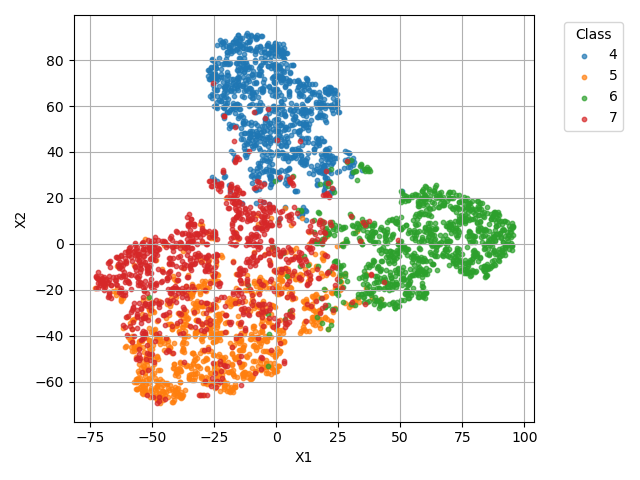}

           \caption{After 3rd task}
    \end{subfigure}
     \begin{subfigure}{0.32\textwidth}
        \centering
           \includegraphics[width=\textwidth]{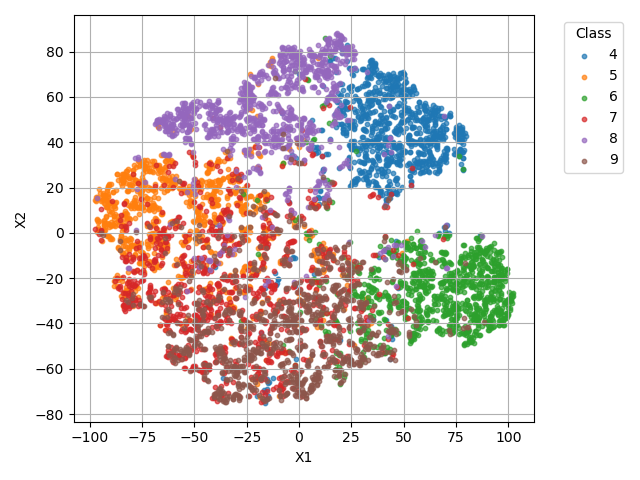}

           \caption{After 4th task}
    \end{subfigure}
     \begin{subfigure}{0.32\textwidth}
        \centering
           \includegraphics[width=\textwidth]{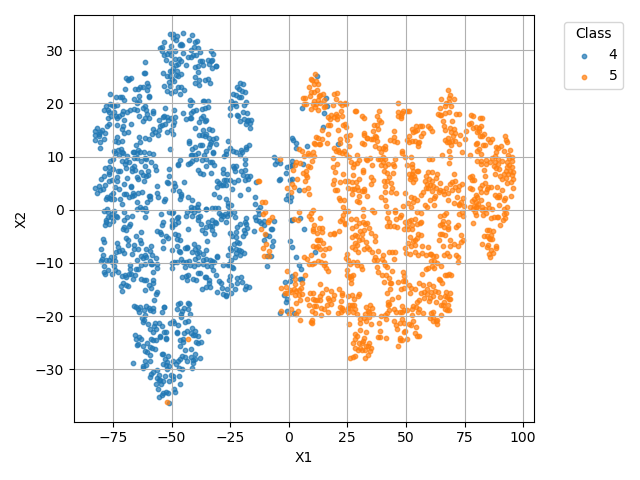}

         \caption{After 2nd task}
    \end{subfigure}
     \begin{subfigure}{0.32\textwidth}
        \centering
           \includegraphics[width=\textwidth]{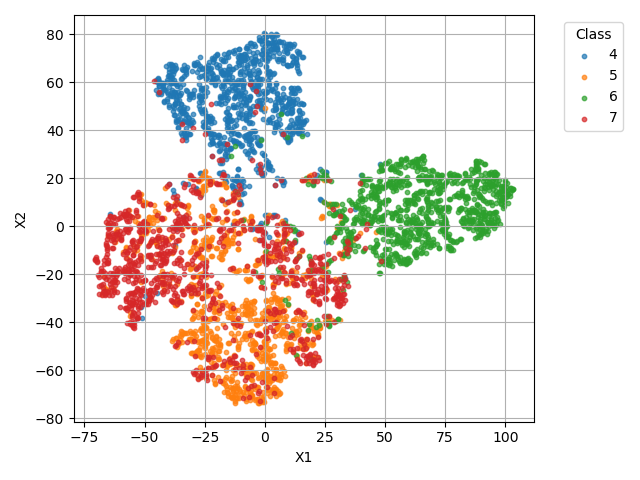}

          \caption{After 3rd task}
    \end{subfigure}
     \begin{subfigure}{0.32\textwidth}
        \centering
           \includegraphics[width=\textwidth]{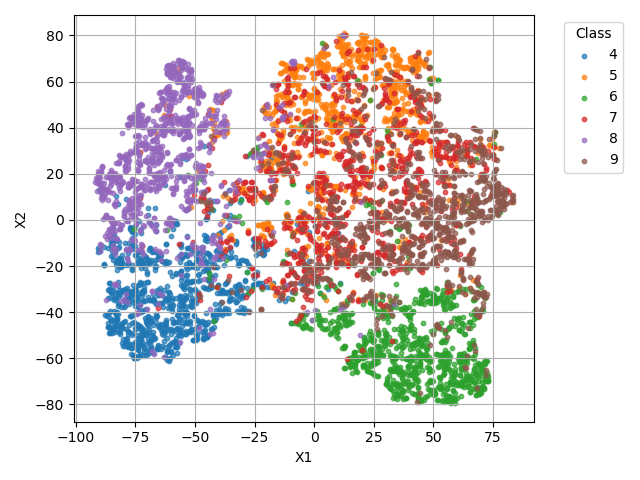}

          \caption{After 4th task}
    \end{subfigure}
    \caption{\textbf{t-SNE visualization of the model output space on clean test data.} 1) when second task is clean (upper row, (a)-(c)) 2) when second task is poisoned using \Bait~attack (bottom row, (d)-(f)).} 
    \label{fig:tsne1}
\end{figure}

\begin{figure}[bt]
    \centering
     \begin{subfigure}{0.35\textwidth}
        \centering
           \includegraphics[width=\textwidth]{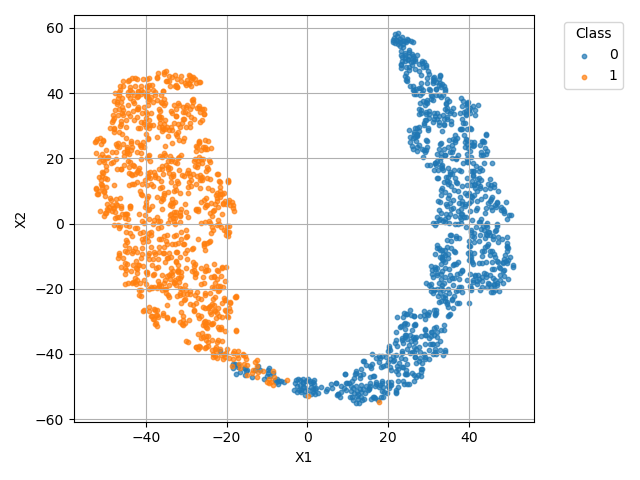}
    \end{subfigure}
     \begin{subfigure}{0.35\textwidth}
        \centering
           \includegraphics[width=\textwidth]{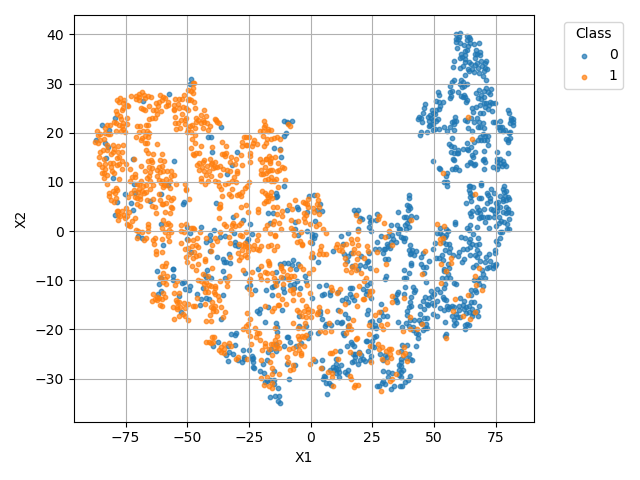}
    \end{subfigure}

      \begin{subfigure}{0.35\textwidth}
        \centering
           \includegraphics[width=\textwidth]{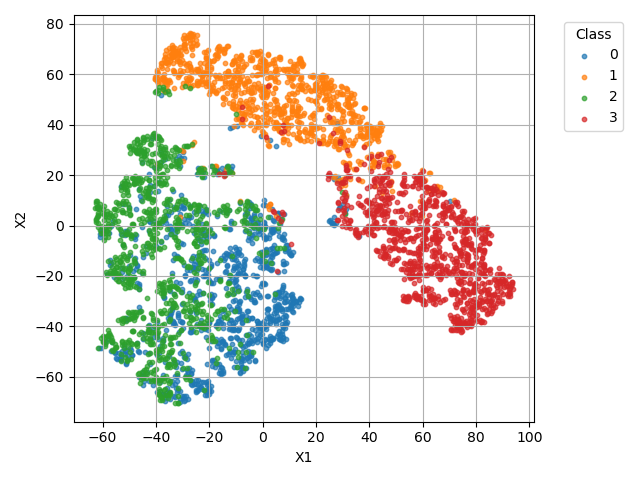}
    \end{subfigure}
     \begin{subfigure}{0.35\textwidth}
        \centering
           \includegraphics[width=\textwidth]{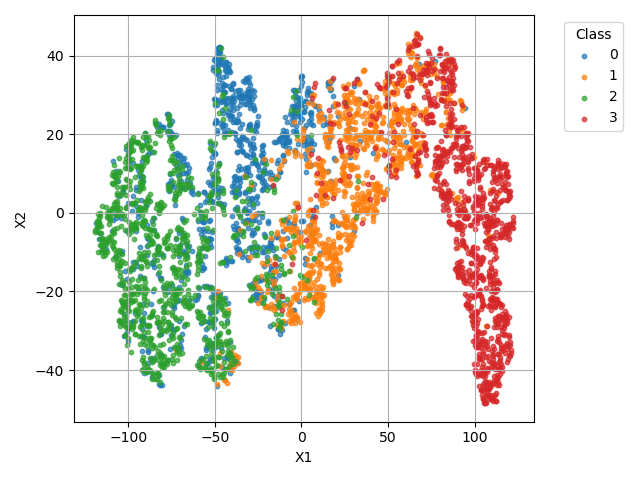}
    \end{subfigure}

      \begin{subfigure}{0.35\textwidth}
        \centering
           \includegraphics[width=\textwidth]{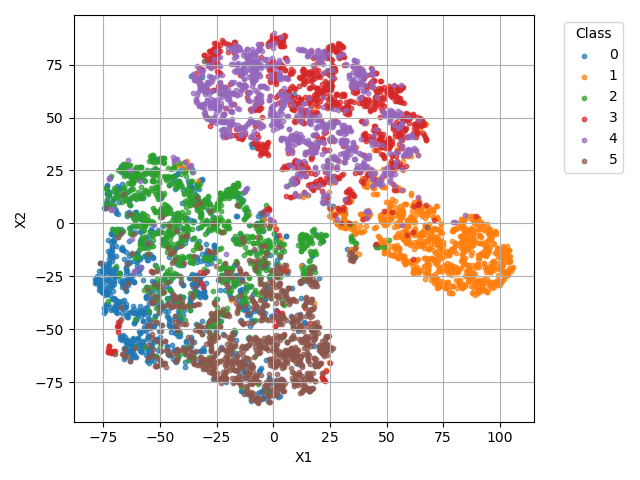}
    \end{subfigure}
     \begin{subfigure}{0.35\textwidth}
        \centering
           \includegraphics[width=\textwidth]{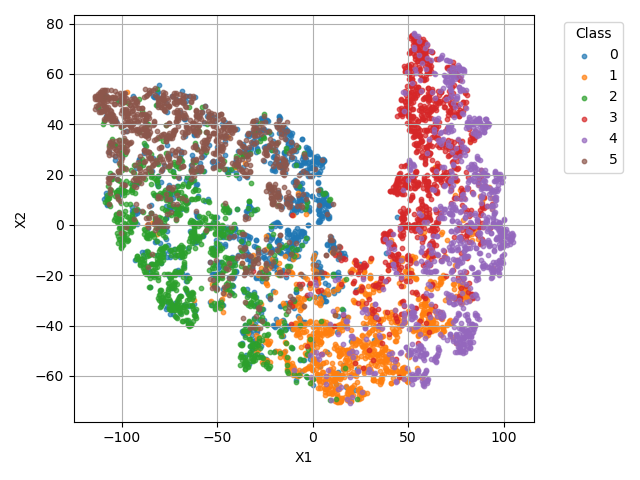}
    \end{subfigure}

      \begin{subfigure}{0.35\textwidth}
        \centering
           \includegraphics[width=\textwidth]{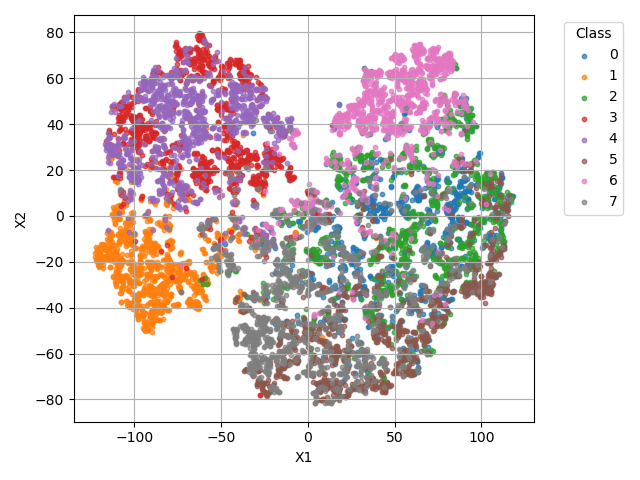}
    \end{subfigure}
     \begin{subfigure}{0.35\textwidth}
        \centering
           \includegraphics[width=\textwidth]{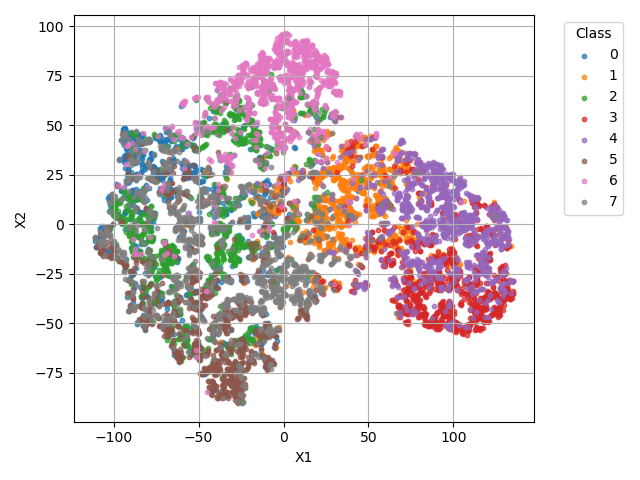}
    \end{subfigure}

          \begin{subfigure}{0.35\textwidth}
        \centering
           \includegraphics[width=\textwidth]{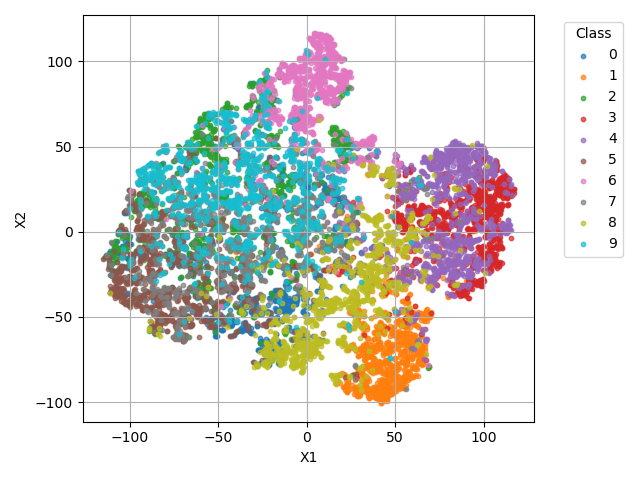}
    \end{subfigure}
     \begin{subfigure}{0.35\textwidth}
        \centering
           \includegraphics[width=\textwidth]{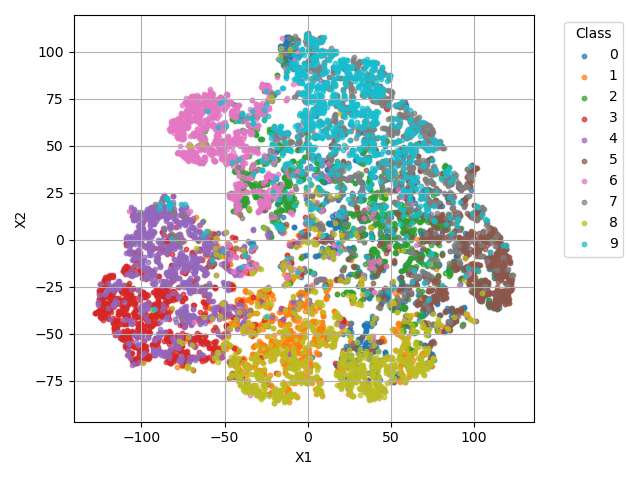}
    \end{subfigure}

    \caption{\textbf{t-SNE visualization of the model output space on clean test data: 1) when all tasks are clean (left) 2) when the first task is poisoned using \Bait~attack (right).}}
    \label{fig:t0_tsne}
\end{figure}

\minisection{Model representation does not collapse after poisoning: t-SNE visualizations.}
We reject the hypothesis that the model collapses to a single representation following the poisoned task. Instead, the~\Bait attack introduces a subtle distribution shift that causes misclassification of clean examples. This is visible when comparing (a) and (d) in Figure\ref{fig:tsne1}, or the first row in Figure~\ref{fig:t0_tsne}. These initial misalignments propagate into future tasks, slightly distorting class boundaries compared to the clean setup.

\section{Experimental Setup details}\label{setup_details}

\minisection{General comments.}
We use FACIL framework~(\cite{masana2022class}) to conduct CL experiments for all methods.
We report results averaged between five runs with different random seeds. 
All experiments and evaluations were performed with Nvidia RTX A5000 GPUs. 

\minisection{Corruptions}
Figure~\ref{fig:STP-corruptions} shows a set of CIFAR-C based corruptions used to poison data in this work. While CIFAR-C dataset is usually used for test-time adaptation scenarios and thus benchmark consists of corrupted test data, we use those predefined transformations to create corrupted images for training the poisoned task. For all results with single poison reported in tables we use $gaussian\_blur$ corruption (see Figure~\ref{fig:STP-corruptions-severities}). For attacks with higher number of poisons (\Multibase~and \Multibait) we use five poisons: $gaussian\_blur$, $gaussian\_noise$, $contrast$, $pixelate$, $jpeg\_compression$. We show attacks using all kinds of corruptions (each as a single poison for \Base~or \Bait ~attacks) on Figure~\ref{fig:base-scatter}.

\minisection{CL setup.} For additional experiments on Tiny Imagenet we split data into six tasks, with 100 classes in the first task, and 20 classes in each following task. Consequently, during our experiments poisoned task ($T_p$) consist of 2 classes for CIFAR10, ten classes for CIFAR100 and 20 classes for Tiny Imagenet. 

\minisection{Models.} Resnet32  (used for CIFAR10/100 experiments) and Resnet18 (used for Tiny Imagenet experiments) implementations are taken from FACIL.

\minisection{Additional parameters.}
We use lambda=10 (distillation loss parameter) for both LWM and LwF methods. For additional experiments on LwM (see Table~\ref{tab:lwm_results}) we use additionally gamma=20 (attention loss parameter). We use data augmentations from FeTrIL~(\cite{petit2023fetrilfeaturetranslationexemplarfree}).  

\minisection{Defense evaluation.} Defense evaluation is done using CIFAR100 dataset. For threshold selection procedure we use second task (to fulfill the requirement of task in the beginning of the stream with the same number of classes as potential poisoned tasks). For poisoned tasks we use 4th task to be consistent with attack evaluation section. As clean tasks we use all 10-class tasks from CIFAR100 splits using multiple random seeds.

\section{Broaden impact}
Our research reveals CL vulnerabilities to data poisoning attacks. To address possible negative impact and minimize the risk of misusing our work we investigate and discuss possible defensive measures. We also present a proposition for defense framework for CL with poison task detection method. Finally, our investigations using STP framework aim to facilitate closing a gap existing between the number of proposed attacks and defenses. 
\end{document}